\documentclass[draft]{agujournal2019}
\usepackage{cmap}
\usepackage[utf8]{inputenc}
\usepackage{rotating} 
\usepackage{booktabs}
\usepackage{amsmath,amssymb}
\usepackage{xcolor}
\usepackage{array}
\usepackage{enumitem}
\usepackage{placeins}
\usepackage[printonlyused,withpage,nolist]{acronym}
\usepackage{graphicx}
\usepackage{newunicodechar}
\newunicodechar{−}{-}
\usepackage{epstopdf}
\usepackage{stfloats}
\usepackage{subfigure}
\usepackage{svg}
\usepackage{tabularx}
\usepackage{booktabs} 
\usepackage{color}
\usepackage{soul}

\usepackage{xr}

\usepackage{natbib}

\usepackage[colorlinks=true,
            linkcolor=blue,
            citecolor=blue,
            urlcolor=blue,
            pdftitle={Your Paper Title},
            pdfauthor={Your Name},
            bookmarksnumbered=true]{hyperref}
\usepackage{array}
\newcolumntype{C}[1]{>{\centering\arraybackslash}p{#1}}

\draftfalse

\journalname{Journal of Geophysical Research: Solid Earth}

\begin{document}
\justifying

\makeatletter
\let\oldNAT@citex\NAT@citex
\renewcommand{\NAT@citex}[3]{%
  \ifNAT@numbers\else\ifNAT@swa\NAT@@open\else\fi
  \if*#2*\else\NAT@cmt#2\fi
  \if\relax\NAT@date\relax\else\NAT@@close\fi\fi
  \begingroup
  \let\NAT@nm\NAT@nmfmt
  \def\NAT@citeauthoryear##1##2##3{%
    \ifnum\NAT@ctype=\z@
      \if\relax##1\relax\else\NAT@name{##1}\fi
      \if\relax##2\relax\else\NAT@spacechar\NAT@name{##2}\fi
      \if\relax##3\relax\else\NAT@spacechar\NAT@name{##3}\fi
    \else
      \ifnum\NAT@ctype=\@ne
        \if\relax##1\relax\else\NAT@name{##1}\fi
        \if\relax##2\relax\else\NAT@spacechar et al.\fi
      \else
        \if\relax##1\relax\else\NAT@name{##1}\fi
        \if\relax##2\relax\else\NAT@spacechar et al.\fi
      \fi
    \fi
    \if\relax\NAT@date\relax\else\NAT@spacechar(\NAT@date)\fi}%
  \oldNAT@citex{#1}{#2}{#3}%
  \endgroup}
\makeatother

\title{ Crustal Structure Imaging of Ghana from \\ Single-Station Ambient Noise Autocorrelations and \\
 Earthquake Arrival Time Inversion}

\authors{
Courage K. Letsa\affil{3},
Hamzeh Mohammadigheymasi\affil{1,2},
Nasrin Tavakolizadeh\affil{4},
Zamir Khurshid\affil{5},
S. Mostafa Mousavi\affil{1},
Martin Schimmel\affil{7},
Cyril D. Boateng\affil{3,6},
and Paulina Amponsah\affil{8} 
}

\affiliation{1}{Department of Earth and Planetary Sciences, Harvard University, 20 Oxford Street, Cambridge, MA 02138, USA}

\affiliation{2}{Atmosphere and Ocean Research Institute, The University of Tokyo, 5-1-5 Kashiwanoha, Kashiwa, Chiba 277-8564, Japan}

\affiliation{3}{Department of Physics, Kwame Nkrumah University of Science and Technology (KNUST), Kumasi, Ghana}

\affiliation{4}{Department of Computer Science, University of Beira Interior, 6201-001 Covilhã, Portugal}

\affiliation{5}{School of Geosciences and Info-Physics, Central South University, 410083 Changsha, China}

\affiliation{6}{West Africa Geophysics Lab, Kwame Nkrumah University of Science and Technology (KNUST), Kumasi, Ghana}

\affiliation{7}{Geosciences Barcelona, GEO3BCN-CSIC, C/ Lluís Solé i Sabarís s/n, 08028 Barcelona, Spain}

\affiliation{8}{Ghana Geological Survey Authority (GGSA), Accra, Ghana}

\correspondingauthor{Hamzeh Mohammadigheymasi}{hmohammadigheymasi@fas.harvard.edu}

\begin{keypoints}
\item Ambient Noise Autocorrelations 
\item Crustal Structure Imaging 
\item Local Earthquake Arrival Time Inversion
\item Seismic interferometry
\item Ghana
\end{keypoints}

\begin{abstract}
The crustal architecture of southern Ghana remains inadequately resolved despite its tectonic significance and resource potential. Existing geological and geophysical studies provide only broad constraints on crustal composition, lacking the resolution to accurately define sediment--basement interfaces or intra-crustal stratigraphy. To address these limitations, we employ single-station ambient noise autocorrelation (SSANA) on continuous waveform data from the Ghana Digital Seismic Network (GHDSN). We extract P-wave reflectivity responses using a processing sequence that involves data pre-processing, Phase Cross-Correlation (PCC) for robust noise correlation, and phase-weighted stacking (PWS) of the derived autocorrelograms. This procedure yields a two-way travel-time (TWT) function representing the zero-offset P-wave reflection response beneath each station, enabling high-resolution imaging of the stratified crustal column. To facilitate depth conversion, we develop an enhanced one-dimensional crustal velocity model for the region. Using a compiled dataset of local earthquake P- and S-wave arrival times from the GHDSN and an additional station in C\^{o}te d'Ivoire, we perform a joint inversion via a grid-search algorithm to derive a regional 1D velocity structure. Our results provide new constraints on the depth and configuration of the Paleozoic basement beneath the Voltaian Basin, demonstrating the efficacy of ambient noise autocorrelation for crustal imaging in sparsely instrumented regions. We also present an updated seismicity catalog, relocated using the new velocity model, and analyze the spatial clustering of seismicity in southern Ghana. This study highlights the utility of passive seismic methods for elucidating crustal structure and evaluating resources in intraplate West Africa and analogous Precambrian terrains.
\end{abstract}

\section*{Plain Language Summary}
Understanding the structure of the Earth's crust is important for reconstructing geological history and identifying natural resources. In southern Ghana, detailed knowledge of the crustal layers has been limited because of the sparse seismic network. In this study, we applied a cost-effective method called ambient noise autocorrelation to image the subsurface. Instead of using earthquakes or artificial explosions, this technique relies on the continuous background vibrations of the Earth (ambient noise) recorded at individual seismic stations. These vibrations are processed and transformed into signals that act like echoes from deep underground rock layers. 
By analyzing these echoes, we identified boundaries between different rock formations. To estimate the depths of these boundaries more accurately, we also developed a new seismic velocity model for the crust based on data from local earthquakes. Our results provide a clearer image of the deep geological structure beneath southern Ghana, including the depth of the ancient basement rocks. This study shows that meaningful information about the Earth's crust can be obtained even in regions with limited seismic coverage, supporting future geological investigations and resource exploration.

\section{Introduction}

Ghana, located in the \ac{WAC}, comprises diverse geological terrains dominated by Paleoproterozoic to Neoproterozoic basement rocks \citep{pasyanos2007top, begg2009lithospheric}. These terrains are underlain by heterogeneous geological formations, including the Paleoproterozoic Birimian Supergroup and the overlying Voltaian Basin and Paleozoic to recent sediments. The crustal structure of southern Ghana remains poorly constrained despite this well-established geological framework, due to the scarcity of high-resolution seismological and geophysical data \citep{GhanaSeismicity2023, AsareBediako2024a}. Previous investigations have attempted to make optimal use of the limited geophysical and borehole datasets available; however, these efforts have yielded only limited and spatially uneven constraints on basement depth and composition.

Within this context of limited crustal-scale constraints, the geometry of the sediment–basement interface has been more extensively studied in the Voltaian Basin using potential geophysical data. Early syntheses \citep{Ako1985} established the basic stratigraphy of the basin, documenting sedimentary successions up to 5--7~km thick. Subsequent airborne geophysical campaigns, notably the BGS/Fugro Ghana Airborne Geophysics Project \citep{BGS2008}, provided regional aeromagnetic and gravity datasets that enabled mapping of basement lineaments and relief, while also confirming sedimentary thicknesses on the order of $\sim$5~km.

Previous information on Moho depth beneath Ghana has been derived primarily from regional gravity studies \citep{ako1985margin, haas2021two, baranov2023new}, seismological investigations \citep{bacon1981earthquake}, and continental to global crustal models \cite{Mooney1998crust, Tedla2011, TUGUME2013250, laske2013update}. These studies report Moho depths of approximately 38–42 km for southern Ghana \citep{bacon1981earthquake, ako1985margin}, while broader-scale models suggest crustal thicknesses ranging from 30 to 45 km. At a broader lithospheric scale, the regional study of \citet{globig2016new} applied an integrated analysis of African crustal and lithospheric mantle structure using elevation, geoid anomalies, and thermal modeling. 
Their results indicate that the West African Craton is underlain by a thick, cold, and mechanically strong lithospheric mantle, with lithospheric thickness locally exceeding $\sim$200 km, and that its topography is primarily supported by lithospheric mantle properties rather than crustal thickness. Toward the southern craton margin, including Ghana, the lithosphere progressively thins and becomes thermally weakened toward the Atlantic margin, consistent with Mesozoic rifting and passive margin formation, highlighting the role of inherited lithospheric structure in controlling deformation, rift localization, and margin evolution in West Africa. In the Keta Basin \citet{Daniel2020} employed 3-D Euler deconvolution and 2-D gravity inversion, constraining basement depth and relief and identifying structural segmentation along the margin. Further offshore, seismic reflection and potential-field analyses by \cite{Antobreh2009} documented the rift–shear margin architecture, offering insights into crustal thinning and basement configuration seaward of the coastal basins.

Seismic studies in Ghana can be broadly classified into two sets: (1) studies conducted using sparse, temporary broadband seismic networks deployed across the region, and (2) seismic imaging studies based on high-resolution \ac{MCS} surveys conducted in the 1970s, mainly focused on offshore Ghana with the primary objective of hydrocarbon exploration. These two approaches differ fundamentally in data coverage, resolution, and imaging capability. Studies based on sparse seismic networks effectively preclude the application of high-resolution imaging techniques, such as body-wave or ambient noise tomography using dense seismic arrays \citep{zhu2000moho}. Consequently, seismological investigations have mostly relied on single-station methods, such as receiver function analysis and Rayleigh-wave ellipticity inversion \citep{osotuyi2025crustal}, which utilize the low-frequency content of teleseismic waves ($<0.5$~Hz). While these approaches provide foundational crustal parameters, such as bulk crustal properties (e.g., crustal thickness and $V_p/V_s$ ratio), they offer a limited-resolution view of the region’s structure \citep{PASYANOS200727, pasyanos2014litho1, Gangopadhyay2007} and predominantly one-dimensional information beneath individual stations.

On this basis, using data from six broadband seismic stations, \citet{akpan2016crustal} analyzed teleseismic earthquakes (M~$\geq$~5.5) recorded between 2009 and 2014 to investigate the crustal structure beneath Ghana. The methodology combined P-wave receiver function analysis, $H$--$\kappa$ stacking, and joint inversion with Rayleigh-wave group velocities. This approach yielded point estimates of crustal thickness in the range of 40--45~km, $V_p/V_s$ ratios between 1.67 and 1.74, and shear-wave velocities ranging from 4.0 to 4.3~km~s$^{-1}$ across Ghana’s distinct tectonic regions. Using a joint inversion of teleseismic P-wave receiver functions, receiver-function horizontal-to-vertical ratios, and Rayleigh-wave ellipticity, \citet{osotuyi2025crustal} determined the crustal structure beneath Ghana. The study presents refined estimates of the crustal thickness (H), the bulk crustal V\textsubscript{P}/V\textsubscript{S} ratio and the shear-wave velocity (V\textsubscript{S}) models beneath two seismic stations in southern Ghana. Their results reveal a general northward crustal thinning from the Gulf of Guinea coast ( $\lesssim$ 44 km) toward the Mauritanian Belt ( $\gtrsim$ 16 km). At station AF.KUKU in southern Ghana, crustal thickness estimates of approximately 48 km are reported.

On the other hand, \ac{MCS} surveys have produced high-resolution seismic depth images mainly in offshore southern Ghana \citep{delteil1974continental, Antobreh2009}. More recently, based on a combined analysis of a detailed grid of 2710~km of multichannel seismic (MCS) lines and potential field data, \citet{Antobreh2009} provide new insights into the structural architecture and tectonic development of the Ghana margin. The results are presented as high-resolution 2D seismic depth sections predominantly perpendicular to the southern margin of Ghana, from the coast to the deep offshore. The study shows that the Ghana continental margin evolved through a combination of extensional rifting in the west and strike-slip–dominated transform deformation in the east, producing strong along-strike variations in crustal structure and basin architecture. Crustal thickness varies significantly along the margin, with the rift-dominated western domain characterized by relatively thick continental crust that thins abruptly toward the transform-dominated eastern domain, where crustal blocks are segmented and structurally complex. Despite the significant achievements of this study, the results are limited to offshore southern Ghana, and the imaged sections are largely restricted to shallow structures due to the depth limitations of controlled-source \ac{MCS} surveys, as well as the strong attenuation of high-frequency seismic energy at greater depths.

These limitations underscore the need to develop or adapt methodologies capable of extracting enhanced structural constraints from the limited data available in sparsely instrumented regions. In this context, passive seismic approaches and in particular \ac{SSANA} provide a robust alternative for imaging crustal structure in data-sparse regions such as southern Ghana. By exploiting continuous ambient seismic noise, \ac{SSANA} retrieves P-wave reflection responses without the need for controlled sources or borehole information, making it both cost-effective and non-invasive. Previous studies have demonstrated the capability of this method to resolve major crustal boundaries and sediment–basement interfaces, even with sparse station coverage \citep{Gorbatov2013,romero2018mapping,taylor2016crustal,benjumea2024subsurface,riahi2026non} among others. Building on these advances, we apply ambient noise autocorrelation to continuous recordings from six broadband stations of the \ac{GHDSN} \citep{gh}. This dataset has recently gained attention within the seismological community due to its high signal quality and its potential for imaging crustal processes \citep{GhanaSeismicity2023,ipiml, mohammadigheymasi2023data,almeida2025numerical}. To enhance the extraction of coherent body-wave reflections, we employ the \ac{PCC} technique \citep{Schimmel2011,Dangwal2021}, which emphasizes phase coherence and improves the stability and resolution of passive images. Subsequently, \ac{PWS} is applied to hourly \ac{PCC}s to further enhance signal coherence and obtain robust daily stacked responses. Data are processed in two frequency bands to resolve structures at different depths: a 3–13 Hz bandpass is used to image shallow crustal features, while a 1–6 Hz bandpass is applied to enhance sensitivity to deeper structures.

The 1D reflectivity series obtained by \ac{SSANA} are TWT time-domain functions, requiring a relatively accurate velocity model for conversion to depth. Existing velocity models for the region lack sufficient accuracy for this purpose \citep{GhanaSeismicity2023}. To address this limitation, we perform a 1D local earthquake arrival-time inversion to derive a regionally representative velocity model tailored to the study area. Phase arrivals reported in a previous study \citep{ipiml} are combined with newly detected phases from a single station located in Côte d’Ivoire. The inclusion of this additional station improves azimuthal coverage for earthquakes predominantly occurring in the southern part of the region, thereby enhancing earthquake detection and the robustness of the velocity inversion. The resulting velocity model is subsequently used to convert the TWT reflectivity sections into depth sections. Primary horizons are then correlated across stations to define laterally continuous acoustic interfaces within the study area. As an additional objective, we relocate the detected earthquakes by \citet{ipiml} and present an updated seismicity map for southern Ghana, accompanied by a brief discussion of its tectonic implications.

The paper is structured as follows: Section~\ref{sec:geology} reviews the geology of southern Ghana; Section~\ref{datasett} and Section~\ref{Methodologyy} presents the dataset and processing methodology; Section~\ref{resultss} and  Section~\ref{discussionn} presents the results of velocity models and reflections and discusses the results; and Section~\ref{conclusionn} summarizes the conclusions and outlines future research directions.

 \begin{figure*}
 \begin{centering}
 \includegraphics[width=\textwidth]{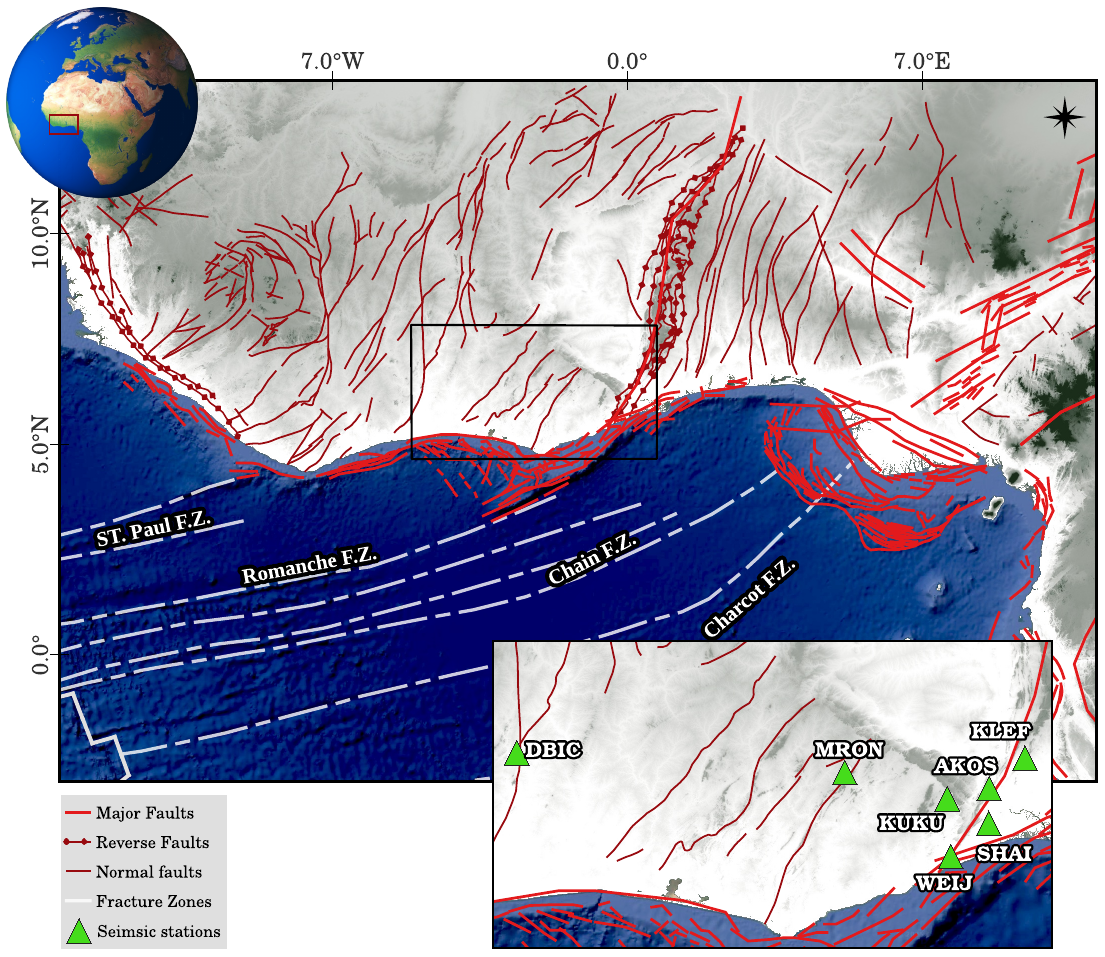}
 \end{centering}
 \vspace{-0.2cm}
\caption{\textcolor{black}{The main panel shows the Gulf of Guinea and its topographic features derived from a Digital Elevation Model (DEM), together with the mapped active faults and fracture zones linking the Mid-Atlantic Ridge to the interior parts of the oceanic ridge system. The study area in southern Ghana is indicated by a black rectangle and is enlarged in the inset at the bottom right. The geographic location of the entire Gulf of Guinea is highlighted by a red rectangle in the inset globe at the top left. Faults in the Gulf of Guinea and East Africa are compiled from \citep{jones2022target}. Reverse, normal, and transform faults are after \citep{thieblemont2016geoafrica}.}}
 \label{GOG3}
 \end{figure*}

\section{Geology and tectonic settings of Ghana}  
 \label{sec:geology}

 Southern Ghana is located in the southern sector of the \ac{WAC} and sits at the junction of two first-order tectonic domains: (i) Paleoproterozoic crust expressed as NE–SW to ENE–WSW-trending Birimian volcanic belts and sedimentary basins, and (ii) the Neoproterozoic–Cambrian Dahomeyide Orogen along the eastern margin, including the Buem and Togo structural units and a high-pressure suture zone \cite{smith2016review}. The region is dominated by crystalline basement rocks of Paleoproterozoic to Neoproterozoic age, which are locally overlain by relatively undeformed sedimentary successions of the Voltaian Basin and other Paleozoic to Recent deposits \citep{pasyanos2007top, begg2009lithospheric, akpan2016crustal}. The Paleoproterozoic Birimian Supergroup in Ghana consists of metavolcanic and metasedimentary sequences forming volcanic belts and intervening basins, with dominant tholeiitic lavas, ultramafic rocks, and subordinate felsic to intermediate volcanic units \cite{smith2016review}.

\begin{figure*}
 \begin{centering}
 \hspace{0.0cm}
 \includegraphics[width=1.1\textwidth]{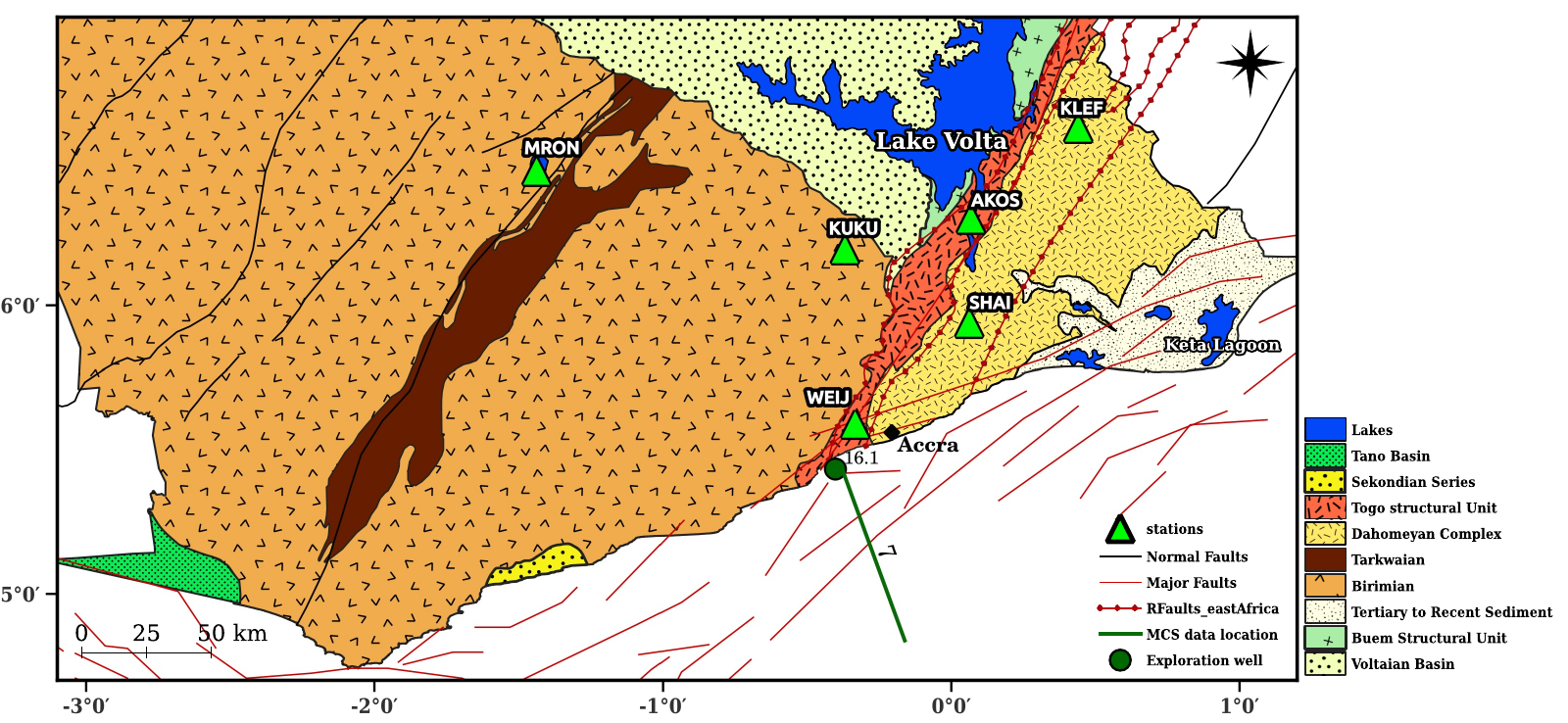}
 \end{centering}
 \vspace{-0.2cm}
\caption{Geological map of Ghana showing the major lithological and structural units across the region (modified from \cite{akpan2016crustal}). Southern Ghana lies within the southern sector of the West African Craton (WAC) and marks the junction between the Paleoproterozoic Birimian terranes and the Neoproterozoic–Cambrian Dahomeyide Orogen along the eastern margin. The legend highlights the principal formations, including Birimian volcanic belts and sedimentary basins, the Tarkwaian and Sekondian basins, the Dahomeyan and Togo Structural Unit, Voltaian sequences, and younger sedimentary cover deposits. The green line labeled 7 represents a seismic profile acquired in 1970 \citep{delteil1974continental}, which is used together with borehole log data from well 16.1 to validate the compiled ambient noise sections in this study.} 	\label{GeologyF}
 \end{figure*}

 The crystalline basement of southern Ghana is mainly composed of the Paleoproterozoic Dahomeyan Complex, the structural units of Togo and Buem, and the Birimian and Tarkwaian Supergroups. Superimposed on this Precambrian basement architecture, the region is characterized by a series of Mesozoic to Cenozoic coastal sedimentary basins, including the Keta, Tano, and Accra basins (Fig.~\eqref{GeologyF}), which developed in response to rifting and faulting associated with the opening of the Atlantic Ocean. The Birimian belts in Ghana are described as long, narrow volcanic/greenstone belts trending approximately NE, separated by intervening metasedimentary basins \cite{smith2016review}. The Tarkwaian Group and its equivalents comprise clastic sedimentary rocks, including sandstones/quartzites, conglomerates, and argillites, commonly preserved in fault-bounded basins within the Birimian belts. In Ghana, this group is best developed in the Ashanti Belt (with a thickness of approximately 2.5 km) \cite{smith2016review}. The Keta Basin forms a graben structure with sedimentary thicknesses of up to 4400 m, consisting of Devonian to Recent nonmarine and marine sediments \citep{Akpati1978GeologicSA}. These basins overlie the Precambrian basement and contribute to the structural and stratigraphic complexity of the region. In addition, the Voltaian Basin is interpreted as a foreland basin, where sediments accumulated in a flexural depression along the margin of the West African Craton, reaching thicknesses of up to $\sim$6,000 m \citep{ako1985margin}.
 
 The \ac{WAC}, particularly along its southern margin, is generally characterized by low levels of seismicity, reflecting its location far from active plate boundaries and low present-day deformation rates ($<$2 mm yr$^{-1}$; \citealp{deprez2013seismic}). However, southern Ghana represents a notable exception, with a documented history of moderate to large earthquakes \citep{ambraseys1986}. The eastern margin of the \ac{WAC} is defined by the Pan-African \ac{AFZ}, which developed during a Neoproterozoic collision between the craton and the Dahomeyan (Lower Precambrian) crust to the east, resulting in the emplacement of Dahomeyan units onto the cratonic margin \citep{ako1985margin}. Despite this intraplate setting, ancient fault systems within the craton remain susceptible to reactivation. 
 
 Two fault systems dominate the seismic hazard narrative in southeastern Ghana: the Coastal Boundary Fault and the \ac{AFZ}. The Coastal Boundary Fault strikes approximately N60–70E and is located about 3–5 km inland from the coast, down-throwing the southern block by several kilometres and forming the northern margin of the Keta Basin. The \ac{AFZ} is a NE–SW-trending fault system extending from west of Accra through Kpong and Ho into Togo and Benin, and is associated with mapped neotectonic normal faults in parts of the Akwapim–Togo belt \cite{ahulu2018probabilistic}. It is further interpreted as reactivation along older thrust boundaries between the Birimian domain and the Togo/Dahomeyan terranes, with evidence of later block-style deformation and development of local normal faults in its southern segment \cite{ahulu2018probabilistic}. Given Ghana’s distance from active plate boundaries, seismic deformation is attributed to two primary mechanisms: stress accumulation associated with the Romanche Fracture Zone, contributing to onshore seismicity in southern Ghana \citep{bacon1981earthquake, kutu2013seismic, Amponsah2014SeismicityAS}, and intraplate tectonic processes operating within the craton and southern Ghana \citep{talwani1991some}. Fault-plane solutions indicate a combination of normal and strike-slip mechanisms \citep{bacon1981earthquake}, and seismicity is concentrated along major inherited structures, particularly at the intersection of the Akwapim and Coastal Boundary fault systems, which is repeatedly identified as a locus of historical and instrumental seismicity \citep{ahulu2018probabilistic, GhanaSeismicity2023, ipiml}. Along the coast, the crust is further influenced by Atlantic opening and transform-margin dynamics. The Coastal Boundary Fault is described as a persistent, high-displacement structure near the coastline, bounding sedimentary troughs that have subsided from Jurassic time to the present and interacting with other faults in the Accra area. This inherited coastal tectonic framework plays a central role in present-day intraplate seismicity patterns in southeastern Ghana \cite{ahulu2018probabilistic}. 
 
 The distribution of the \ac{GHDSN} stations across southern Ghana spans two major crustal domains: the Paleoproterozoic Birimian terrane of the West African Craton and the Neoproterozoic Dahomeyide belt along its southeastern margin. Stations such as WEIJ, SHAI, AKOS, and KLEF are located within the Dahomeyide belt, a Pan-African metamorphosed zone characterized by high-grade gneisses, migmatites, amphibolites, and locally high-pressure mafic granulites (e.g., at Shai), indicating a felsic-to-intermediate upper crust with significant mafic interlayers and a dominantly mafic lower crust \citep{AidooDahomeyideBelt2020, attoh2008tectonic}. In contrast, KUKU and MRON, situated farther inland within the Birimian domain, overlie arc-related Paleoproterozoic crust composed of metavolcanic greenstone belts (mafic protoliths) intruded by abundant granitoids, resulting in an overall intermediate-to-felsic upper crust underlain by a more mafic lower crust typical of accreted arc terranes \citep{EISENLOHR1992313}.

 Receiver-function analyses for southern Ghana support a predominantly felsic-to-intermediate bulk crust (Poisson’s ratio $\sim$0.24–0.26), consistent with granitoid-rich upper crustal sections overlying denser mafic lower crust \citep{akpan2016crustal, osotuyi2025crustal}. Receiver-function and joint inversion studies further provide quantitative constraints on crustal thickness, with a regional joint inversion across western Africa reporting H–k estimates for stations near the Gulf of Guinea, including AF.KUKU = 43.6 $\pm$ 2.47 km and AF.SHAI = 35.2 $\pm$ 3.56 km. These stations are described as metasedimentary site conditions and exhibit associated Vp/Vs ratios of 1.74 $\pm$ 0.028 and 1.85 $\pm$ 0.072, respectively \citep{osotuyi2025crustal}. 
 The receiver-function study of \citet{akpan2016crustal} indicates that the Birimian, Dahomeyan terrain, and Togo Structural Unit exhibit broadly similar crustal structures, with an average Moho depth of $\sim$44 km and Poisson’s ratios of $\sim$0.24–0.26, interpreted as a bulk felsic–intermediate composition. It further infers a thick mafic lower crust (12–17 km) beneath the Neoproterozoic terrains. Crustal thickness beneath the Ghanaian seismic stations varies between 41 and 45 km, with shear-wave velocities of 4.0–4.3 km/s defining lower crustal layers of approximately 12 km at SHAI and up to $\sim$17 km at MRON and WEIJ. Collectively, these results reflect a lateral transition from relatively coherent Birimian arc-type crust in the interior to compositionally heterogeneous, tectonically reworked Dahomeyide crust toward the southeastern margin of the craton.

 At the lithospheric scale, this heterogeneous basement–basin architecture is reflected in the deep crustal structure, characterized by a vertically stratified crust comprising a mafic lower crust overlain by a more felsic–intermediate upper to middle crust. Complementary models indicate a broader regional trend of northward crustal thinning, extending from the coastal domain along the Gulf of Guinea toward the Mauritanian Belt, with Moho depths decreasing from $\sim$44 km in the south to values exceeding $\sim$16 km toward the craton margin \citep{osotuyi2025crustal}. Both studies converge on a predominantly felsic to intermediate crustal composition, with individual crustal layers typically 12–17 km thick. Despite the recognized seismic activity, the crustal structure of southern Ghana remains inadequately constrained due to its complex accretionary history and the limited availability of high-resolution geophysical and seismological data \citep{ipiml, AsareBediako2024a}. The intraplate tectonic setting further complicates structural interpretation, as deformation is subtle and poorly expressed at the surface \citep{talwani1991some}. In addition, multiple episodes of magmatism, metamorphism, and sedimentation have produced a heterogeneous lithosphere, limiting the effectiveness of conventional seismic imaging techniques in resolving laterally coherent crustal structures. 
 
 In a broader tectonic context, the offshore Ghana margin provides complementary constraints. Seismic reflection studies classify this margin as a transform continental margin \citep{delteil1974continental, attoh2004seismic, edwards1997synthesis, Antobreh2009}. Seismic profiles document rifting processes within the Gulf of Guinea associated with the separation of southern Ghana from Brazil. Pre-rift sequences consist of Paleozoic sediments overlying Paleoproterozoic basement and include Jurassic volcanic intrusions. Syn-rift strata of Early Cretaceous age record an initial phase of oblique rifting, while a mid-Cretaceous unconformity marks the transition to post-rift sedimentation. Post-rift sequences, spanning the Late Cretaceous to Holocene, were deposited during passive margin development \citep{Antobreh2009}.

 To place onshore observations in a broader tectonic context, the offshore Ghana margin provides important complementary constraints. Seismic reflection studies classify this margin as a transform continental margin \citep{delteil1974continental, attoh2004seismic, edwards1997synthesis, Antobreh2009}. Seismic profiles document rifting processes within the Gulf of Guinea associated with the separation of southern Ghana from Brazil. Pre-rift sequences consist of Paleozoic sediments overlying Paleoproterozoic basement and include Jurassic volcanic intrusions. Syn-rift strata of Early Cretaceous age record an initial phase of oblique rifting, while a mid-Cretaceous unconformity marks the transition to post-rift sedimentation. Post-rift sequences, spanning the Late Cretaceous to Holocene, were deposited during passive margin development \citep{Antobreh2009}.

Here, we provide a discussion on the  interpretation of the geological interfaces revealed by the ambient noise sections in different stations. Considering the location of stations on the regions's geological formations across the eastern margin of the West African Craton (WAC),  MRON and KUKU are located in a similar formation within the interior of the Birimian Supergroup, composed of Paleoproterozoic ($\sim$2.2--2.0~Ga) metavolcanic, metasedimentary, and granitoid assemblages formed during the Eburnean orogeny. WEIJ lies in the border of the eastern margin of the Birimian domain, adjacent to the Togo Structural Unit, a Neoproterozoic passive-margin sequence thrust westward during the Pan-African orogeny, but very close to the Birimian Supergroup.

\section{Dataset} \label{datasett}

We processed continuous broadband seismic data recorded at six stations of the \ac{GHDSN} network \citep{gh}
 (Fig.~\ref{GOG3}), deployed in southern Ghana and spanning the period from October 2012 to March 2014. This dataset has been previously analyzed for local earthquake detection and 1-D crustal velocity inversion in southern Ghana \citep{ custodio2022seismicity, GhanaSeismicity2023, ipiml, mohammadigheymasi2023data}. In addition, we incorporate waveform data from a single broadband station located in Côte d’Ivoire, operated by the \ac{GTSN} \citep{https://doi.org/10.7914/sn/gt}, which provided seismic monitoring over the same time interval. All waveform data were obtained from the \href{http://ds.iris.edu/ds/nodes/dmc/data/types/waveform-data/}{\ac{IRIS}}
 Data Management Center (DMC). All stations are three-component broadband instruments, with a sampling rate of 100~Hz for the \ac{GHDSN} stations and 40~Hz for the \ac{GTSN} station. The detailed processing steps and methodological framework are described in the following subsections.

\section{Methodology} \label{Methodologyy}

This study employs two principal methodological approaches: (1) retrieval of the Earth's reflection response as a function of \ac{TWT} using single-station seismic interferometry, and (2) development of a regionally constrained 1-D velocity model to accurately convert the \ac{TWT} reflection sections into the depth domain. The latter provides the essential velocity constraints for interpreting the reflection images.

\subsection{Reflection Response Estimation via Seismic Interferometry} \label{REF_RES}

We apply the single-station autocorrelation of ambient noise (\ac{SSANA}) method introduced by \cite{Schimmel2011}, following the general processing scheme of \cite{romero2018mapping} and optimizing key parameters for our study area. Vertical-component seismograms are used, as they are most effective for isolating and imaging P‑wave reflection energy from subsurface interfaces directly beneath each station. The data is derived from single-station noise recordings to obtain a dataset primarily sensitive to crustal structure beneath each station. The workflow comprises three main stages: (i) Preprocessing, in which continuous waveform data are segmented into 1‑day records, instrument response and baseline trends are removed, followed by band‑pass and notch filtering, and each daily record is subdivided into 1‑hour time windows; (ii) Autocorrelation, where auto‑correlation functions (\ac{ACFs}) are computed for each preprocessed 1‑hour noise window to retrieve coherent seismic wavefield information in the form of a \ac{TWT} reflection series approximating the zero‑offset impulse response beneath the station; and (iii) Stacking, in which the hourly \ac{ACFs} are temporally stacked to enhance the signal‑to‑noise ratio and stabilize coherent reflection phases.

\begin{figure*}
	\begin{centering}
				\includegraphics[width=\textwidth]{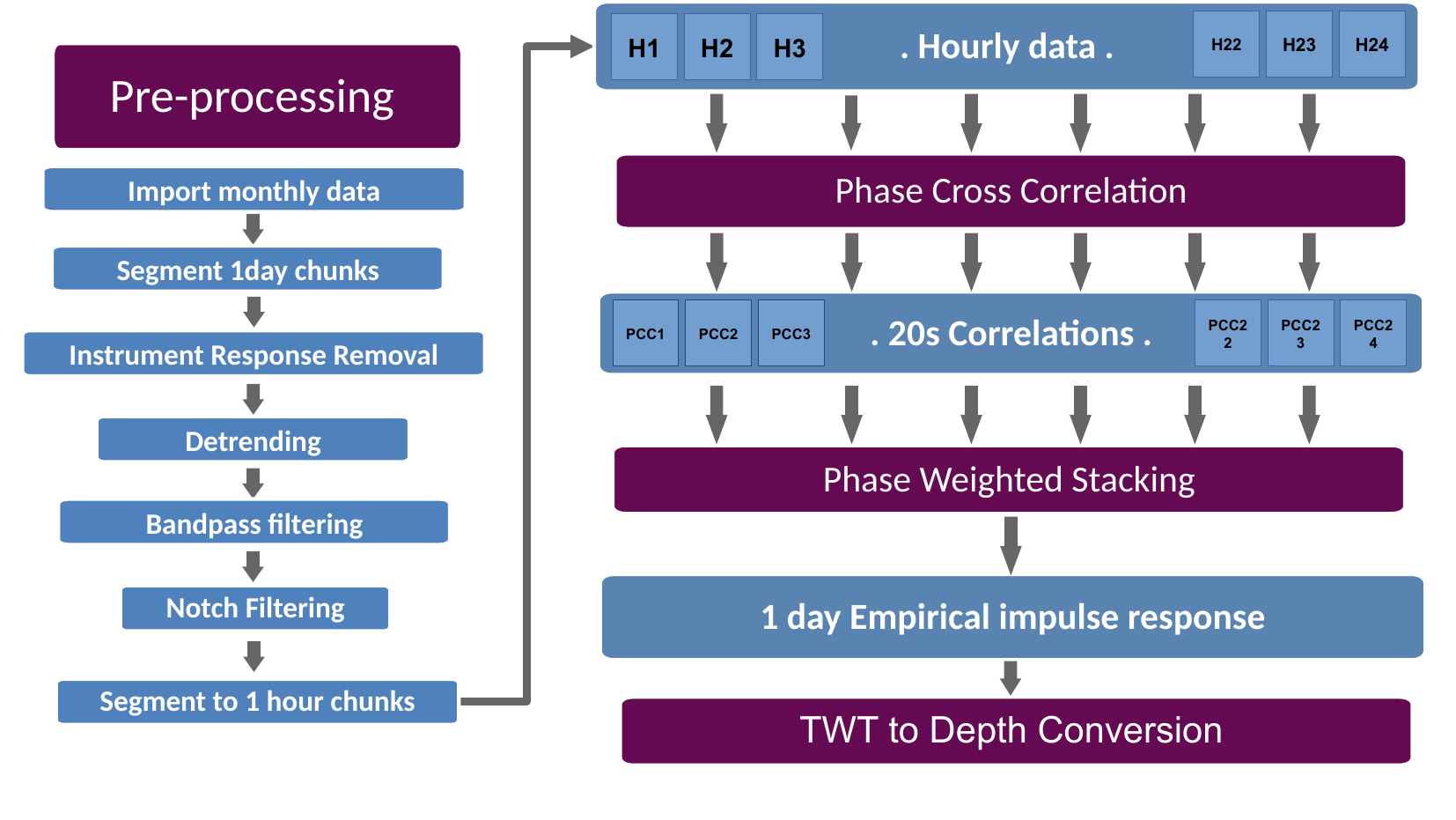}
				\vspace{0.3cm}
		\includegraphics[width=\textwidth]{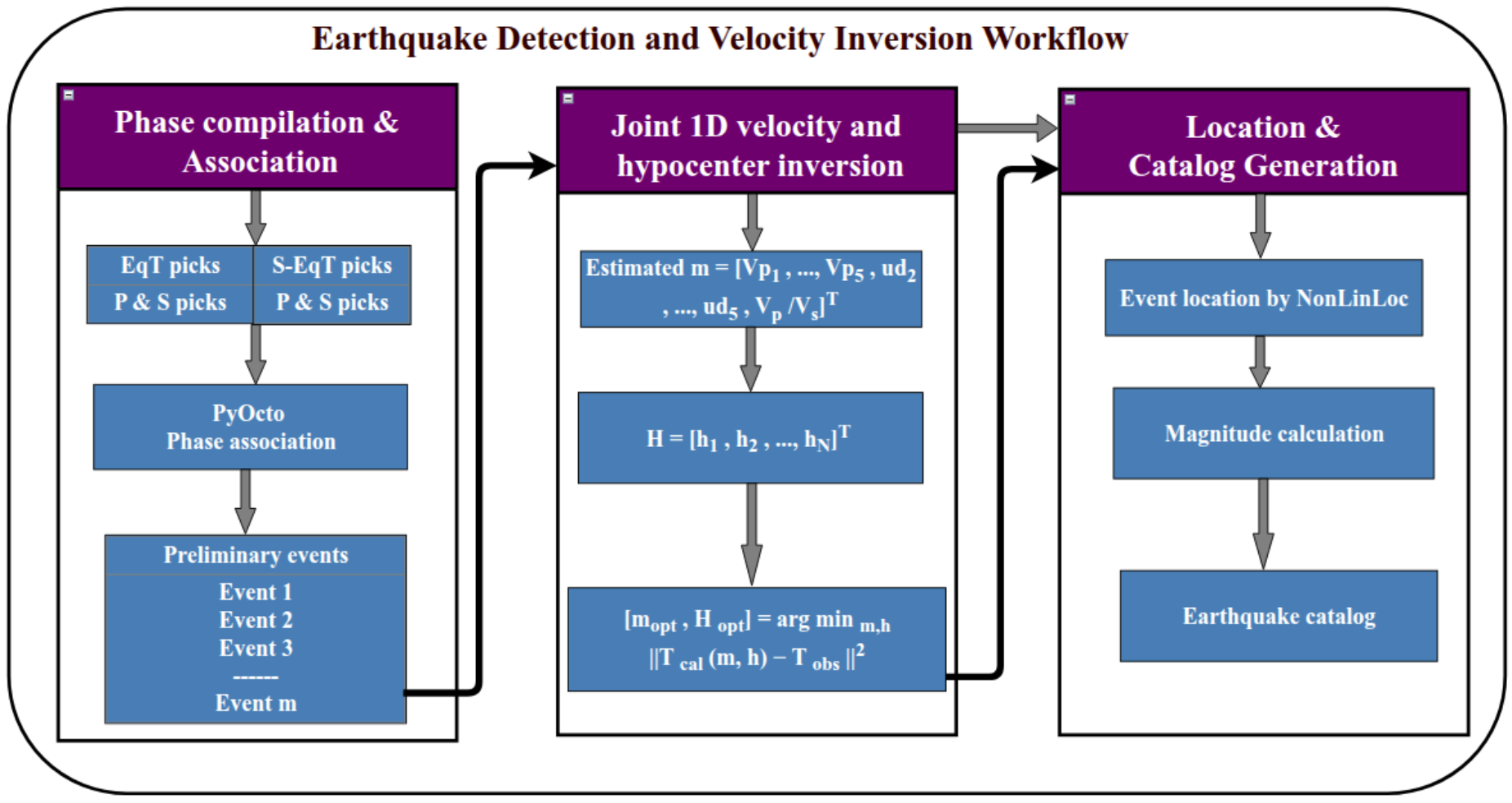}	
	\end{centering}
	\vspace{-0.2cm}
	\caption{
\textbf{Top:} Workflow for ambient seismic noise reflection series estimation. Continuous seismic records are first preprocessed, and the data are then segmented into one-hour windows. Cross-correlations are computed up to 20-s lags. These functions are subsequently stacked using phase-weighted stacking to generate the daily reflection series. Finally, the results are converted from two-way travel time (TWT) to depth using the velocity model estimated in the bottom workflow.
\textbf{Bottom:} Earthquake detection and velocity inversion workflow. P and S phases from continuous waveforms are detected using the EQT and SEQT models. The phases are then associated using the PyOcto method, yielding a set of preliminary events to be used as input for the joint inversion, which simultaneously estimates hypocentral parameters and a parametric 1D velocity model. The optimal velocity model is subsequently employed for TWT to depth conversion and final earthquake relocation of detected events. 
	}
	\label{PCCworkflow}
\end{figure*}

The daily autocorrelation functions are used to identify time variability as might be caused by varying weather or other noise source conditions and to assess the robustness of the reﬂection response.

\subsubsection{Pre-processing}

The processing workflow begins with the segmentation of continuous waveform data into 1-day chunks, which are subsequently divided into twenty-four 1-hour segments. For each segment, long-period trends and baseline drifts are removed by fitting a 5th-order spline with 1000 samples. This detrending preserves shorter-period ambient noise energy generated within the crust.

A critical component of our methodology is the strategic selection of frequency bands to resolve crustal structures at different scales. Higher frequencies are necessary for imaging shallow, thin layers and minimizing interference from the zero-lag autocorrelation peak \citep{benjumea2024subsurface, romero2018mapping,nacif2024insights,esquivel2025mapping,ruigrok2012global}, while lower frequencies are required for deeper imaging due to attenuation and larger layer thicknesses. Consequently, we apply zero-phase Butterworth filters with passbands of 1–6 Hz for deep structures and 3–13 Hz for shallow features.

Following filtering, the instrument response is deconvolved using StationXML metadata, stabilized by a prefilter with corner frequencies at 0.3, 0.5, 13.0, and 16.0 Hz. Residual long-period artifacts are removed by a second detrending step using a 3rd-order spline with 1000 control points, and a time-domain taper is applied to prevent edge artifacts during autocorrelation.

For station AKOS, a persistent resonance at $\sim$6 Hz required an additional adaptive notch filter to suppress this monochromatic noise. Finally, the processed 1-hour segments for both frequency bands were aggregated for autocorrelation computation with a maximum time lag of 20 s.

\subsubsection{Phase Autocorrelation}
Several methods exist for retrieving seismic reflections from ambient noise, most notably Cross-Correlation of Geometrically Normalized noise (CCGN) \cite{bensen2007processing} and Phase Cross-Correlation (PCC) \cite{Schimmel2011}. While CCGN mitigates amplitude fluctuations through energy normalization, it remains sensitive to high-amplitude transients, often necessitating temporal or spectral whitening \citep{bensen2007processing,schimmel2018low}. Extending the analysis to the time–frequency domain, \citet{schimmel2018low} demonstrated the effect and importance of time- and frequency-domain normalizations using low-frequency autocorrelations.

In contrast, PCC is amplitude-independent, relying solely on instantaneous phase similarity derived from the analytic signal. The discrete phase autocorrelation at lag $k$ is given by:

\begin{equation}
C_{\text{PCC}}[k] = \frac{1}{N} \sum_{n=0}^{N-1-k} \frac{s[n] s[n+k]}{|s[n]| |s[n+k]|}
\end{equation}

where $s[n]$ is the analytic signal and $|s[n]|$ is its envelope. By measuring waveform similarity through phase alignment, PCC efficiently suppresses high-amplitude transients without amplitude normalization. Given its superior performance in imaging deeper crustal structures and robustness under low signal-to-noise conditions \citep{romero2018mapping}, we adopt PCC for this study.

\subsubsection{Phase Weighted Stacking}

To obtain a single, high signal-to-noise ratio correlation function for each day, the hourly PCC traces are stacked using phase-weighted stacking (\ac{PWS}). This technique enhances coherent arrivals by weighting the conventional linear stack with the phase coherence calculated across all input traces. The discrete phase-weighted stack, $S_{\text{PW}}[n]$, is computed as:

\begin{equation}
S_{\text{PW}}[n] = S_{\text{linear}}[n] \times \left| \frac{1}{N} \sum_{i=1}^{N} e^{j\phi_i[n]} \right|^p
\end{equation}

where $S_{\text{linear}}[n] = \frac{1}{N} \sum_{i=1}^{N} s_i[n]$ represents the linear stack at the discrete time index $n$, $N$ denotes the number of traces, $\phi_i[n]$ is the instantaneous phase of the $i$-th trace, and $p$ is the weighting exponent/order. A phase coherence value approaching unity signifies strong phase alignment, while smaller values mark less coherent signals, thereby preferentially accepting coherent arrivals, such as reflections, while effectively suppressing incoherent noise. Based on tests conducted on our dataset, an exponent of $p = 3$ was selected to optimally alleviate noise and enhance deeper crustal reflections.

The stacking process incorporates tolerances for minor timing discrepancies and variations in trace length among the hourly correlations, ensuring a robust daily estimate of the empirical impulse response. The method is founded on the principle that instantaneous phase coherence serves as a powerful metric for extracting signals from noise cross-correlations \cite{Schimmel2011}.  A summary of the complete processing workflow is presented in Figure \eqref{PCCworkflow}.

 \FloatBarrier

\subsection{Velocity model estimation for TWT to Depth Conversion}

The stacked daily correlation functions form the basis for structural interpretation. The resulting 1-D autocorrelation function approximates a zero-offset vertical reflection response, imaging acoustic impedance contrasts associated with crustal interfaces. However, the zero-offset geometry inherent to the autocorrelation method precludes direct estimation of interval velocities through moveout analysis \citep{claerbout1968synthesis}. Therefore, converting the measured \ac{TWT} to depth requires an independent velocity model as a constraint.

Early constraints on crustal velocities in Ghana were based on uniform layer approximations. Initial studies by \citet{bacon1979recent} assumed constant velocities of $V_p = 5.6$~km/s and $V_s = 3.3$~km/s. Subsequently, \citet{bacon1981earthquake} refined this by proposing a depth-dependent 1-D model with a linear velocity gradient based on analysis from a temporary seismic array:
\begin{equation} \label{velocity}
    V_p(z) = V_{p0} + \alpha z,
\end{equation}
where $z$ is depth in kilometers, $V_{p0} = 5.6$~km/s is the near-surface velocity, and $\alpha = 0.04$~s$^{-1}$ is the velocity gradient. This model, and variations thereof using nearly constant velocities (e.g., $V_p \approx 5.55$~km/s; \citet{akoto1992monitoring, amponsah2002seismic}), have been widely adopted for regional seismicity studies.

A more recent attempt to estimate a velocity model for the region was conducted by \citet{GhanaSeismicity2023}. However, the robustness of their results is limited by the small number of earthquakes incorporated into the inversion (73 events) and the short epicentral distances of the dataset, which reduced sensitivity to deeper crustal structure.

To overcome these limitations and provide robust velocity constraints for depth conversion in this study, we derived a new 1-D crustal velocity model for the region through local earthquake tomography. The model derivation followed a two-stage workflow: first, compiling an enhanced earthquake catalog from local stations; and second, performing a 1-D velocity inversion using the refined arrival-time dataset.

\subsubsection{Phase data compilation and association} \label{asso}
We compiled an enhanced local earthquake P- and S-wave arrival-time dataset from two primary sources: (1) previously published phase picks from the \ac{GHDSN} as reported by \citet{ipiml}, and (2) reprocessed continuous waveform data from station DBIC, part of the \ac{GTSN}, located in Côte d’Ivoire near the border with Ghana. The DBIC data were processed using the deep learning (DL) workflow introduced by \citet{ipiml}. The DeepScan phase detection method employed here has been effectively applied to high-resolution seismicity analysis in diverse geographic settings \citep{khurshid2025comprehensive, carvalho2025application}. The inclusion of station DBIC significantly improved azimuthal coverage—particularly for hypocenters concentrated in southern Ghana—and increased the number of high-quality phase picks. This expanded dataset provides tighter constraints on the deeper portions of the crustal velocity structure, thereby enhancing the robustness of time-to-depth conversions for the interpreted reflection horizons.

The compiled phase candidates were then associated into seismic events using the DL–based PyOcto phase association framework \citep{munchmeyer2023pyocto}, which is designed to robustly link picks in low signal-to-noise environments. For a detection to be accepted as a valid seismic event, we required that the association include at least 2 station with both P- and S-wave arrivals and a minimum of 5 total picks across the network. These criteria reduce false detections while retaining sensitivity to small-magnitude seismicity. During association, PyOcto was configured to constrain event searches to a defined spatial volume spanning 0–13°N latitude, 10°W–7°E longitude, and 0–45~km depth, covering southern Ghana and adjacent areas of Côte d’Ivoire and Togo. This geographic and depth limitation prevents linking picks from distant regional or teleseismic phases and ensures that the resulting solutions reflect local crustal seismicity.

\subsubsection{1D velocity inversion} \label{joint}

Using the associated phases, we applied a joint inversion method to simultaneously estimate hypocenter parameters for all events and 1D velocity model parameters. We used the HYP program from the SEISAN earthquake analysis software package \citep{lienert1995computer,havskov1999seisan}, employing a grid-search method to comprehensively explore the velocity model parameter space by varying layer upper depths ($u_d$), layer P-wave velocities ($V_p$), and the $V_p/V_s$ ratio.

A 6-layer velocity model was defined. By fixing the upper depths of the first layer ($u_{d1}=0$ km) and the sixth layer ($u_{d8}=45$ km), we estimated 10 model parameters: $\boldsymbol{m}=[V_{p1},...,V_{p7},u_{d2},...,u_{d7},V_p/V_s]^T \in \mathbb{R}^{14}$.

In this joint inversion scheme, the optimized velocity model, $\boldsymbol{m}_{opt}$, and the hypocentral parameters, $\boldsymbol{H}=[\boldsymbol{h}_1,\boldsymbol{h}_2,... , \boldsymbol{h}_N]^T$ (where $\boldsymbol{h}i=[x_i,y_i,z_i, t{0i}]$ for $i=1,...,N$ with $N = 453$ recorded earthquakes), are calculated by minimizing the RMS residual between the observed and calculated arrival times:

\begin{equation} \label{eqq} [\boldsymbol{m}{opt}, \boldsymbol{H}{opt}] =\arg \min_{\boldsymbol{m},\boldsymbol{H}} {||\boldsymbol{T}{cal}(\boldsymbol{m},\boldsymbol{H})-\boldsymbol{T}{obs}||^2_2}. \end{equation}

In Eq.\eqref{eqq}, $\boldsymbol{T}{obs}=[t{1}, t_{2}, ... , t_{M}]^T \in \mathbb{R}^{M}$ is the vector of observed arrival times for $M=3734$ picked phases (see Sec. \ref{asso}). The calculated arrival times, $\boldsymbol{T}_{cal}(\boldsymbol{m},\boldsymbol{H})$, are obtained for each hypocenter-station pair by:

\begin{align} \label{ts} t_{cal}(\boldsymbol{m}, \boldsymbol{h})=t_0+\int_{\mathcal{L}} \frac{d\ell}{V(\ell)}, \end{align}

where the integral is taken along the minimum-time ray path, $\mathcal{L}$, from the hypocenter to the receiver, and $d\ell$ is the ray element in the direction of wave propagation \citep{gheymasi2013local}.

A two-step search strategy is implemented. In the first step, a coarse grid search over $V_p$ and $u_d$ is performed, while $V_p/V_s$ is explored over an initial test range of $[1.62, 1.64, \ldots, 1.80]$. This allows identification of a narrow neighborhood around the optimal ratio, constrained to $V_p/V_s \in [1.69, 1.70, 1.71]$. In the second step, a fine-grid search is conducted within this restricted range, jointly exploring $V_p$, $u_d$, and $V_p/V_s$ over the discrete parameter space defined in Table~\ref{ts}. This hierarchical search scheme substantially reduces the computational cost of model-space exploration by constraining the $V_p/V_s$ range prior to the fine-grid inversion. The resulting optimal models are presented in the following section.

\subsubsection{Reflection Response Time--to--Depth Conversion}
The reflection section obtained using the method described in Sec.\ref{REF_RES} displays seismic horizons in the time domain, expressed as two-way travel time (TWT). These horizons approximate the zero-offset impulse response beneath the station. To relate these features to geological structure, we incorporate the depth-interval velocity model estimated in Sec.\ref{joint}. Assuming near-vertical wave propagation, which is appropriate for zero-offset seismic data and receiver-function–based interpretations, we express TWT as a function of depth. We denote TWT by $\tau$ and define the cross-correlation section in the lag-time domain as $C_{\tau}(\tau,n)$, where $n$ represents the day index.

Using the one-dimensional P-wave interval velocity model $V_{\mathrm{p}}(z)$ from Sec.~\ref{joint}, the relationship between time and depth is given by:

\begin{equation} \tau(z) = 2 \int_{0}^{z} \frac{1}{V_{\mathrm{p}}(\zeta)} , d\zeta. \label{eq:tau_integral} \end{equation}

Using Eq.~\eqref{eq:tau_integral}, the lag-time section is reparameterized into the depth domain as $C_{z}(z,n) := C_{\tau}(\tau(z),n)$. In practice, this transformation is implemented by interpolating each daily trace $C_{\tau}(\tau,n)$ from the calculated $\tau(z)$ values onto a uniform depth grid. This approach strictly accounts for vertical velocity gradients, avoiding the biases inherent in constant-velocity approximations, particularly at greater depths.

\subsubsection{ Earthquake Location }

Seismic event locations were determined using the \texttt{NonLinLoc} algorithm \citep{lomax2000probabilistic,Lomax2009}, a widely adopted nonlinear earthquake location package designed to estimate hypocentral parameters within arbitrarily complex velocity structures. \texttt{NonLinLoc} implements a probabilistic grid-search approach based on Bayesian inference, enabling the sampling of the full posterior probability density function (PDF) for the event location, rather than relying on linearized approximations. This framework provides robust uncertainty estimation, which is particularly critical in regions where ray paths are strongly heterogeneous or poorly constrained.

Travel times were computed using the \texttt{Grid2Time} module, which calculates travel-time fields via finite-difference eikonal solvers on a 3-D velocity grid. The use of complete travel-time grids eliminates the necessity for repeated ray tracing during the inversion process and ensures stable solutions for secondary arrivals or complex wavefields. Hypocenter solutions were derived from the maximum likelihood point of the posterior PDF, while the corresponding confidence ellipsoids were extracted from the marginal distributions. This nonlinear, grid-based strategy reduces sensitivity to initial guesses and model nonlinearity, making \texttt{NonLinLoc} particularly suitable for tectonically complex regions, as well as for networks with sparse or irregular station geometries.

\section{Results} \label{resultss}

\subsection{Time-domain reflection response sections}
We constructed two sets of time-domain reflection response sections for stations of the GHDSN network using band-limited microseismic wavefields. The data were separated into two frequency bands: 3–13 Hz, optimized for imaging shallow crustal structures, and 1–6 Hz, targeting deeper crustal interfaces. This frequency partitioning exploits the higher-frequency components of the microseismic spectrum to enhance vertical resolution in the shallow crust, where geological layering is typically thin, while retaining the lower-frequency energy required for coherent imaging of reflections from deeper interfaces  \citep{wapenaar2010tutorial,draganov2008retrieval}. Such a multiband strategy improves depth-dependent resolution by balancing frequency-controlled penetration depth and spatial resolving power.

\begin{figure*}
 \begin{centering}
 \hspace{0.0cm}
 \includegraphics[width=\textwidth]
 {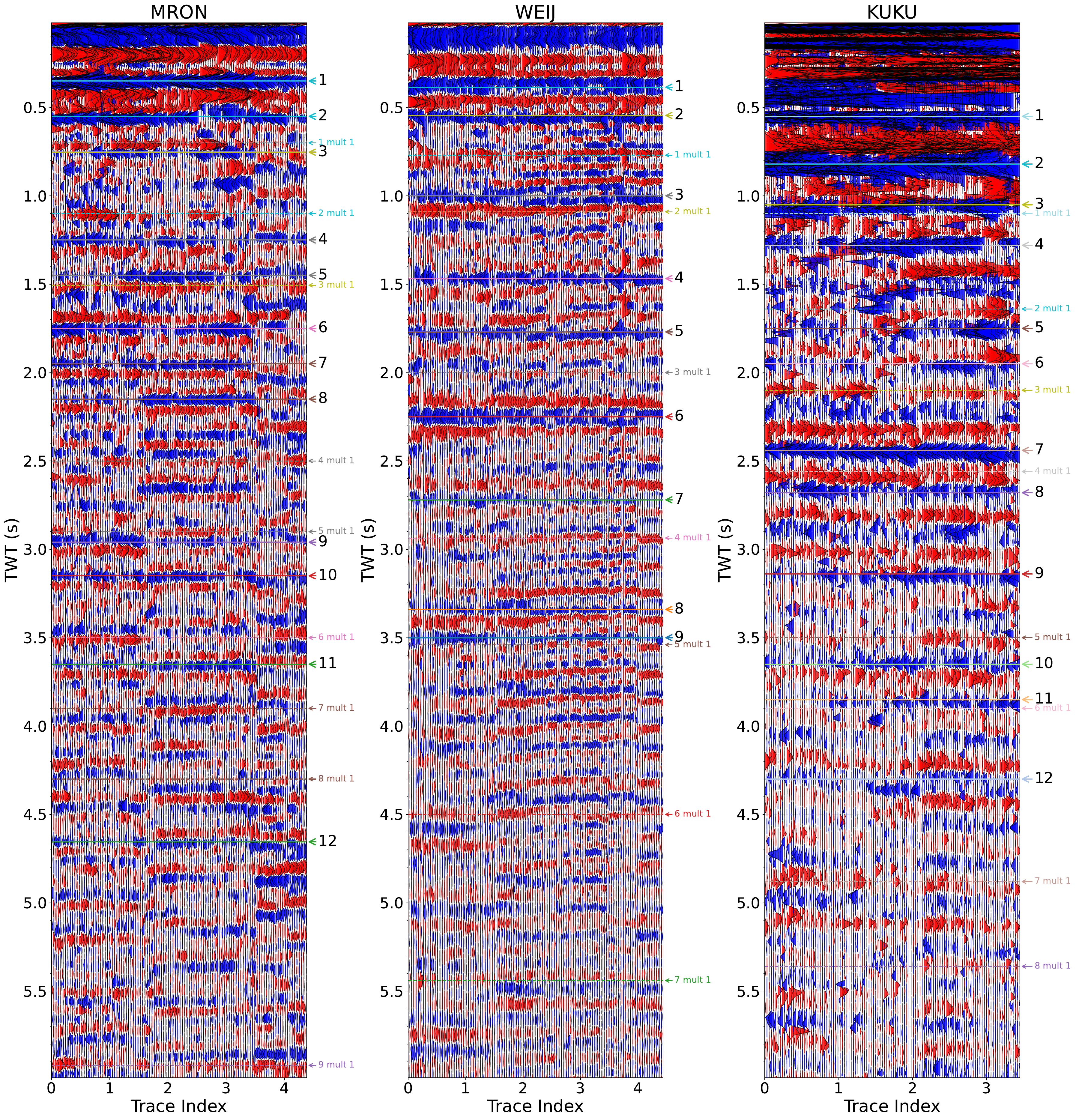}
 \end{centering}
 \vspace{-0.2cm}
 \caption{ High-frequency reflection responses (3–13 Hz) obtained using SAANA, displayed in two-way travel time (TWT), for the stations  MRON, WEIJ, and KUKU, which is optimized for imaging shallower crustal structures. Primary and multiple reflection phases are labeled in the noise-derived sections. Coherent reflection phases appear as laterally continuous horizontal bands, indicating persistent subsurface impedance contrasts retrieved from ambient seismic noise. The enhanced reflectivity confirms the presence of key crustal interfaces beneath the station.}
 \label{shallow_1st}
 \end{figure*}

\begin{figure*}
 \begin{centering}
 \hspace{0.0cm}
 \includegraphics[width=\textwidth]
 {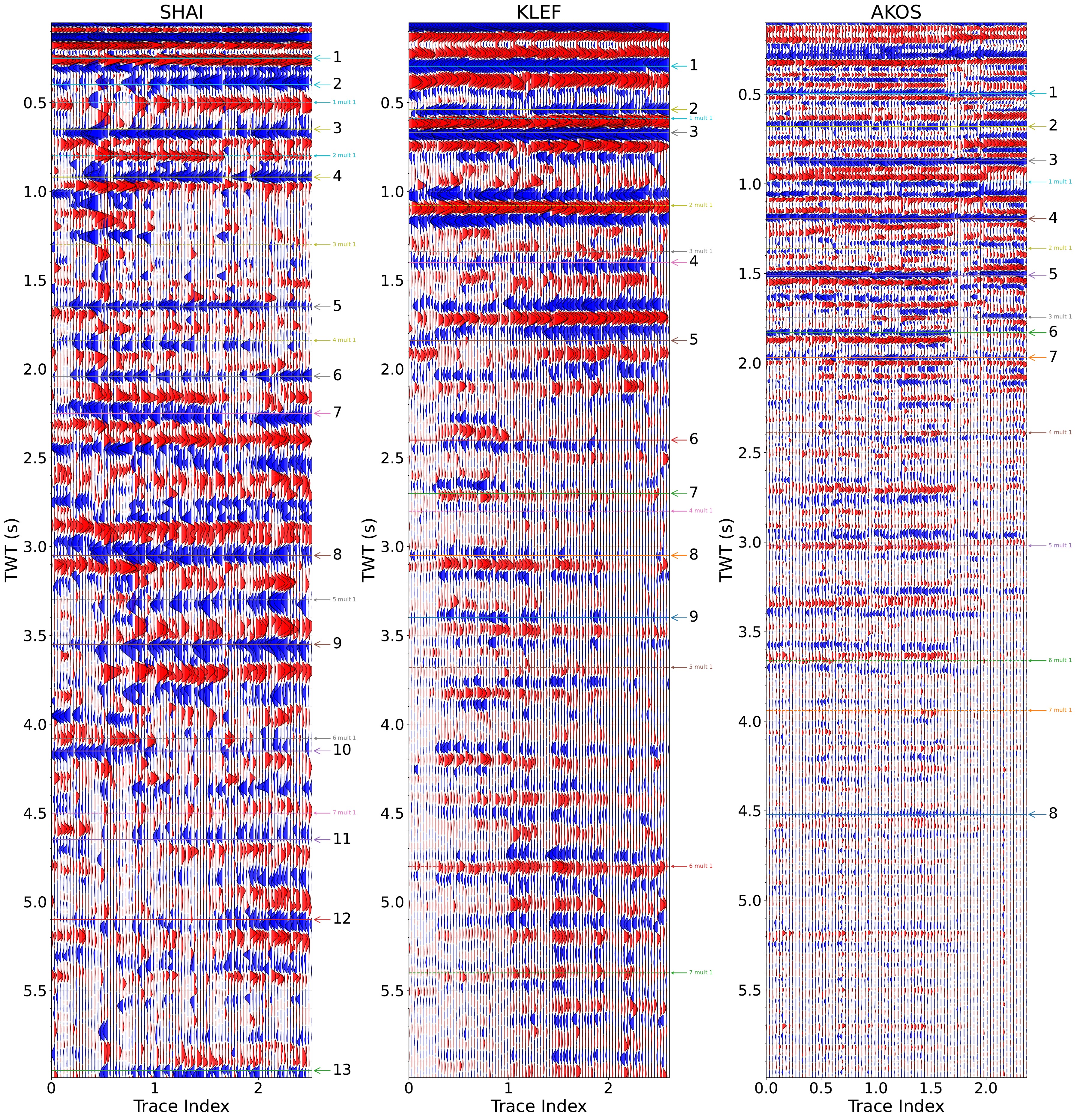}
 \end{centering}
 \vspace{-0.2cm}
 \caption{  A similar high-frequency reflection responses (3–13 Hz) obtained using SAANA, displayed in two-way travel time (TWT), for the stations  SHAI, KLEF, and AKOS. Coherent reflection phases appear as laterally continuous horizontal bands, indicating persistent subsurface impedance contrasts retrieved from ambient seismic noise. The enhanced reflectivity confirms the presence of key crustal interfaces beneath the station.}
 \label{shallow_2nd}
 \end{figure*}

The stacked reflection response sections for each frequency band are further grouped according to two distinct geological domains in southern Ghana, as described in Section~\ref{sec:geology}. The cratonic domain includes stations WEIJ (located near the craton–basin boundary), KUKU, and MRON, which are shown in Fig.~\ref{shallow_1st}. The basin domain comprises stations AKOS, SHAI, and KLEF, illustrated in Fig.~\ref{shallow_2nd}.

Primary reflection arrivals and their multiples were initially identified by integrating their two-way travel times (TWTs), under the assumption that multiples occur at approximately twice the TWT of the corresponding primary reflections. Consistent with the autocorrelation of ambient seismic noise, primary reflections associated with an increase in acoustic impedance across subsurface interfaces exhibit negative polarity (blue-filled amplitudes), whereas their corresponding first multiples display positive polarity (red-filled amplitudes). 

\begin{table}[htbp]
    \centering
    \caption{Crustal thickness estimates for stations KUKU and SHAI in southern Ghana \citep{osotuyi2025crustal}. 
    H(1) represents estimates from forward modeling of Moho Ps arrival times, 
    H(2) from H--k stacking, Average Crustal thickness, and Poisson's ratio. }
    \label{tab:ghana_depth_estimates_simple}

    \small 
    \resizebox{\textwidth}{!}{%
    \begin{tabular}{@{}lC{2.4cm}C{2.4cm}C{2.8cm}C{2.4cm}@{}}
        \toprule
        \textbf{Station} &
        \textbf{Moho Ps (H(1)) (km)} &
        \textbf{H--k Stacking (H(2)) (km)} &
        \textbf{Average Crustal Thickness (km)} &
        \textbf{Poisson's Ratio} \\
        \midrule
        KUKU & 44 $\pm$ 4.0 & 43.6 $\pm$ 2.47 & 45 $\pm$ 3.0 & 0.24 \\
        SHAI & 45 $\pm$ 3.0 & 35.2 $\pm$ 3.56 & 43 $\pm$ 3.0 & 0.25 \\
        \bottomrule
    \end{tabular}%
    } 
    \normalsize 
\end{table}

 \begin{sidewaysfigure}
	\centering
	\includegraphics[width=\textheight]{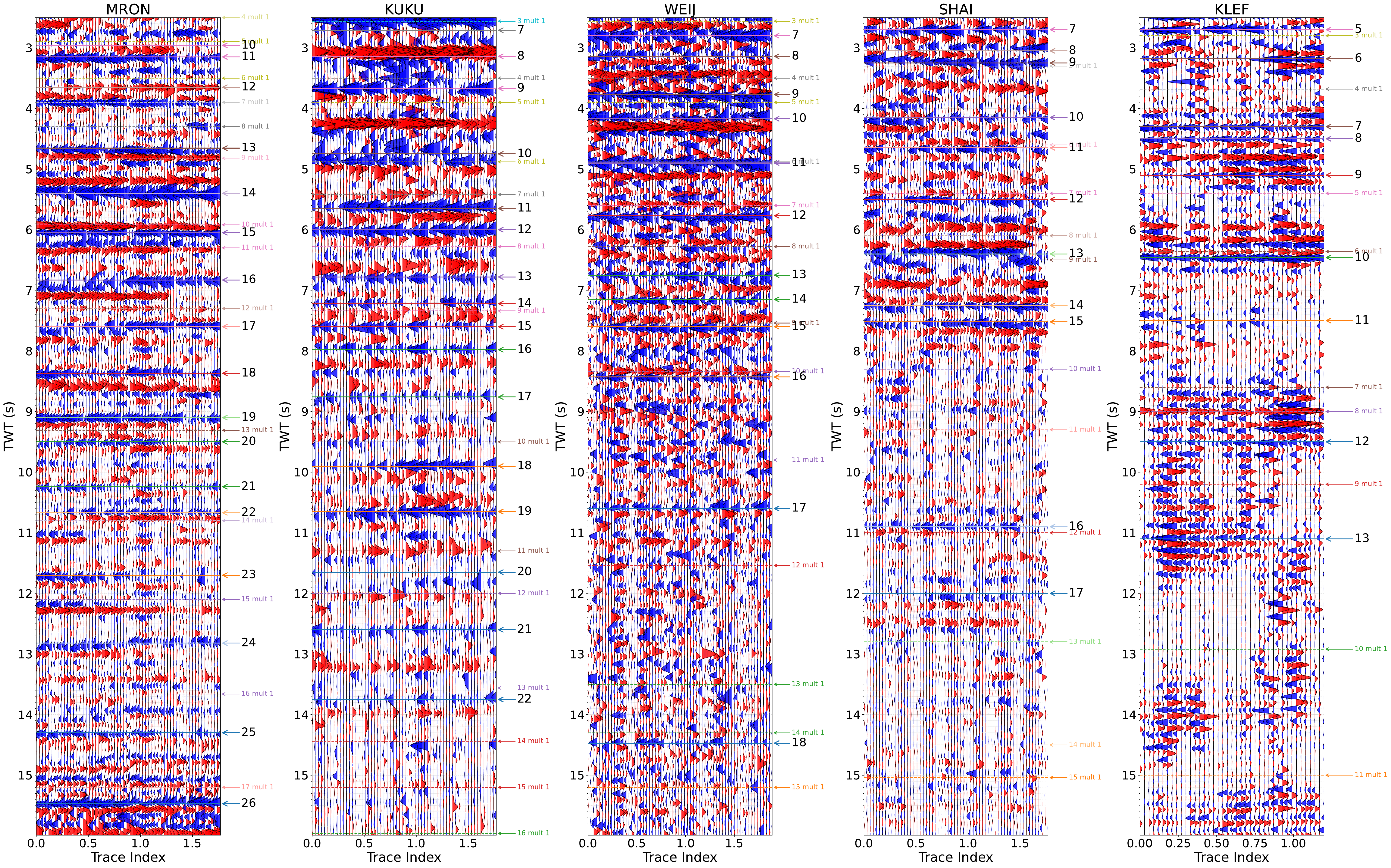}
	\caption{Low-frequency reflection responses obtained using SAANA, displayed in two-way travel time (TWT), for the stations KUKU, MRON, WEIJ, KLEF, and SHAI. The autocorrelation functions were computed in the 1–6 Hz frequency band, which is optimized for imaging deeper crustal structures. Primary and multiple reflection phases are labeled in the noise-derived sections. For the AKOS seismic station, the section is shown as two-way travel time (vertical axis) versus recording days (horizontal axis). Coherent reflection phases appear as laterally continuous horizontal bands, indicating persistent subsurface impedance contrasts retrieved from ambient seismic noise. The enhanced reflectivity confirms the presence of key crustal interfaces beneath the station.}
	\label{fig:akos_landscape}
\end{sidewaysfigure}

\subsection{Estimated 1D velocity} \label{vell}

\begin{figure*}
	\begin{centering}
		\includegraphics[width=\textwidth]{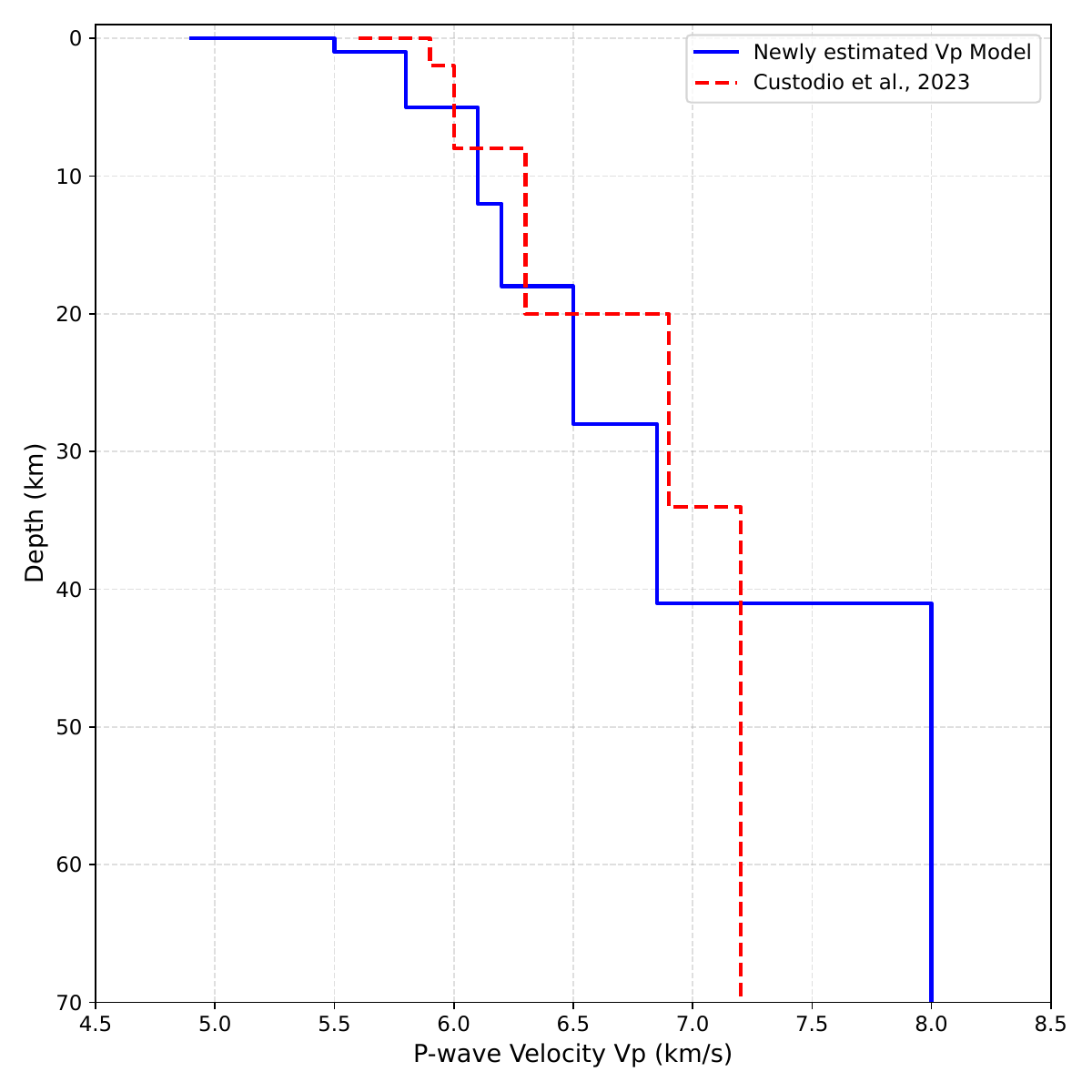}
	\end{centering}
	\caption{\textcolor{black}{Comparison of the newly derived 1D P-wave velocity model for southern Ghana with the earlier model of \cite{custodio2022seismicity}. Both profiles illustrate seismic velocity as a function of depth, highlighting near-surface low-velocity sediments, the transition into crystalline basement, and velocity gradients within the mid- to lower crust. The overlay underscores consistencies and differences between the two models, providing improved constraints on crustal structure beneath the station.}}
	\label{akos_velocity_model}
\end{figure*}

The newly derived one-dimensional P-wave velocity model for Southern Ghana (Table~\ref{final_vel_model}, Figure~\ref{akos_velocity_model}) reveals a progressively increasing velocity structure with depth. The uppermost layer (0--1~km) exhibits a velocity of 4.90~km/s, corresponding to unconsolidated sedimentary cover. This is underlain by a layer extending from 1 to 5~km depth with a velocity of 5.50~km/s. A velocity of 5.80~km/s characterizes the 5--12~km depth range.

Mid-crustal layers show velocities increasing from 6.10~km/s in the 12--18~km interval to 6.20~km/s between 18 and 28~km depth. In the lower crust, velocity increases to 6.50~km/s from 28 to 41~km depth. The layer extending from 41 to 80~km is characterized by a velocity of 6.85~km/s, interpreted as a high-velocity lower crustal layer. A distinct velocity jump to 8.00~km/s is observed at 80~km depth, marking the transition to typical upper mantle velocities (Moho).


Comparison of the newly estimated 1D velocity structure with the \citet{GhanaSeismicity2023} model reveals distinct differences across both depth ranges. In the shallow section, the new model estimates significantly lower P-wave velocities ($V_p$), beginning at 4.90 km/s at the seafloor and increasing to 5.50 km/s at 1.0 km depth, whereas the Custodio 2023 model presents much higher initial values of 5.9 km/s and 6.1 km/s for these respective depths. This trend of lower velocity estimates persists into the deeper parts of the model; for instance, at mid-to-deep crustal levels (approx. 12–18 km), the new model estimates velocities between 6.10 km/s and 6.20 km/s, consistently lower than the 6.3–6.5 km/s range observed in the reference model. Similarly, at greater depths (~41 km), the new model reaches 6.85 km/s, remaining lower than the 7.2 km/s value reported by Custodio (2023) at 45 km.

\begin{table}
	\caption{Final 1D-minimum Velocity structure. The depth of a layer is the starting point from which the seismic P-wave velocity $V_p$ is applied, measured from the seafloor. $V_p/V_s$ ratio was fixed to 1.70 for all layers} \label{final_vel_model}
	\begin{tabular*}{\linewidth}{ @{\extracolsep{\fill}} lc *{13}c @{}}
		\toprule
		\midrule
		\small 
		\small  \textbf{Layer}  & \small  \textbf{Depth (km)} & \small \textbf{ $V_p$(km/s)}  &  \small  \textbf{$V_s$ (km/s)} & \small  \textbf{$V_p/V_s$ }   \\
		\midrule
		\small 1 & \small 0.0  & \small 4.90 & \small 2.88  & \small 1.70 \\
		\small 2 & \small 1.0  & \small 5.50 & \small 3.24  & \small 1.70 \\
		\small 3 & \small 5.0  & \small 5.80 & \small 3.41  & \small 1.70 \\
		\small 4 & \small 12.0 & \small 6.10 & \small 3.59  & \small 1.70 \\
		\small 5 & \small 18.0 & \small 6.20 & \small 3.65  & \small 1.70 \\
		\small 6 & \small 28.0 & \small 6.50 & \small 3.82  & \small 1.70 \\
		\small 7 & \small 41.0 & \small 6.85 & \small 4.03  & \small 1.70 \\
		\small 8 & \small 80.0 & \small 8.00 & \small 4.71  & \small 1.70 \\
		\midrule
		\bottomrule
	\end{tabular*} 
\end{table}

\subsection{High-resolution Seismicity map}

\begin{figure*}
	\begin{centering}
		\hspace{0.0cm}
		\includegraphics[width=\textwidth]
		{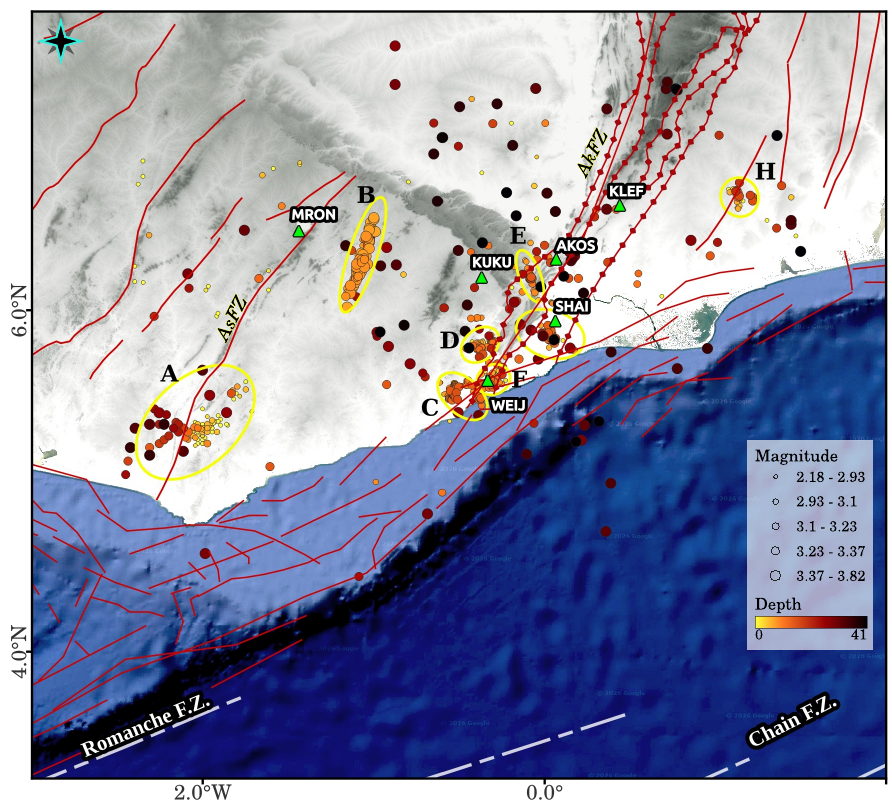}
	\end{centering}
	\vspace{-0.2cm}
	\caption{This map shows the spatial distribution of the compiled earthquake catalog used in this study, comprising 741 events primarily located in southern Ghana and adjacent offshore areas. Seismic clusters A–H, outlined by green ellipses, are displayed together with major mapped tectonic features, including the \ac{AsFZ} and \ac{AkFZ} (black lines) and the offshore \ac{RFZ} (dashed white line). Background topography is draped over the regional bathymetric surface.}
	\label{Final_catalogg}
\end{figure*}

A total of 740 events were detected following our re-processing of the GHDSN data, which incorporated an additional station from the Ivory Coast in Abidjan [see Fig. \ref{Final_catalogg}]. This represents a significant increase, with 1.61 times more events than the 461 reported by \citet{ipiml}. This higher event count is primarily attributed to improved phase association. The spatial distribution of seismicity in the final compiled catalog (Fig.~\ref{Final_catalogg}) corresponds well with previously identified active zones in southern Ghana \citep{amponsah2002seismic,ipiml,musson2014seismicity}. In addition to delineating these known structures, the seismicity pattern reveals previously undocumented features that have been classified into eight main clusters. These clusters align with known active zones in southern Ghana, particularly around the \ac{AkFZ} and the \ac{AsFZ}. Previous studies have similarly shown that seismicity in Ghana is concentrated in the southern part of the country, closely associated with these fault systems \citep{amponsah2002seismic,kutu2013seismic}. The major orientation of the clusters follows the general NE–SW structural trend of the region, connecting inland active faults to the offshore \ac{RFZ}. This geometry supports earlier suggestions that stresses from the Mid-Atlantic Ridge influence inland seismicity through these offshore fracture zones \citep{kutu2013seismic}.

\begin{table}[htbp]
    \centering
    \caption{Summary of crustal thickness estimates for seismological stations in southern Ghana. 
    Data compiled from \cite{akpan2016crustal}, published in \textit{Tectonophysics}~[1--3]. 
    H(1) represents estimates from forward modeling of Moho Ps arrival times, 
    H(2) from H--k stacking, and H(3) from joint inversion~[4]. 
    Mafic lower crust thickness corresponds to depth intervals with 
    shear-wave velocities ($V_s$) between 4.0 and 4.3~km/s~[5].}
    \label{tab:ghana_depth_estimates_simple}

    \small 
    \resizebox{\textwidth}{!}{%
    \begin{tabular}{@{}lC{2.4cm}C{2.4cm}C{2.4cm}C{2.8cm}C{2.4cm}@{}}
        \toprule
        \textbf{Station} &
        \textbf{Moho Ps (H(1)) (km)} &
        \textbf{H--k Stacking (H(2)) (km)} &
        \textbf{Joint Inversion (H(3)) (km)} &
        \textbf{Average Crustal Thickness (km)} &
        \textbf{Poisson's Ratio} \\
        \midrule
        KUKU & 44 $\pm$ 4.0 & 45.0 $\pm$ 2.0 & 45 $\pm$ 2.0 & 45 $\pm$ 3.0 & 0.24 \\
        MRON & 40 $\pm$ 3.0 & 46.0 $\pm$ 3.0 & 42 $\pm$ 2.0 & 43 $\pm$ 3.0 & 0.24 \\
        SHAI & 45 $\pm$ 3.0 & 43.0 $\pm$ 3.0 & 42 $\pm$ 2.0 & 43 $\pm$ 3.0 & 0.25 \\
        AKOS & 46 $\pm$ 3.0 & 45.0 $\pm$ 2.0 & 45 $\pm$ 3.0 & 45 $\pm$ 3.0 & 0.26 \\
        KLEF & 43 $\pm$ 4.0 & 47.0 $\pm$ 3.0 & 42 $\pm$ 2.0 & 44 $\pm$ 3.0 & 0.24 \\
        WEIJ & 42 $\pm$ 4.0 & 40.0 $\pm$ 3.0 & 42 $\pm$ 2.0 & 41 $\pm$ 3.0 & 0.26 \\
        \bottomrule
    \end{tabular}%
    } 
    \normalsize 
\end{table}

\section{Discussion} \label{discussionn}

\subsection{Validation of the Time-domain reflection response sections}

\begin{figure}
	\begin{centering}
		\hspace{-5.2cm}
		\includegraphics[width=1.7\textwidth]{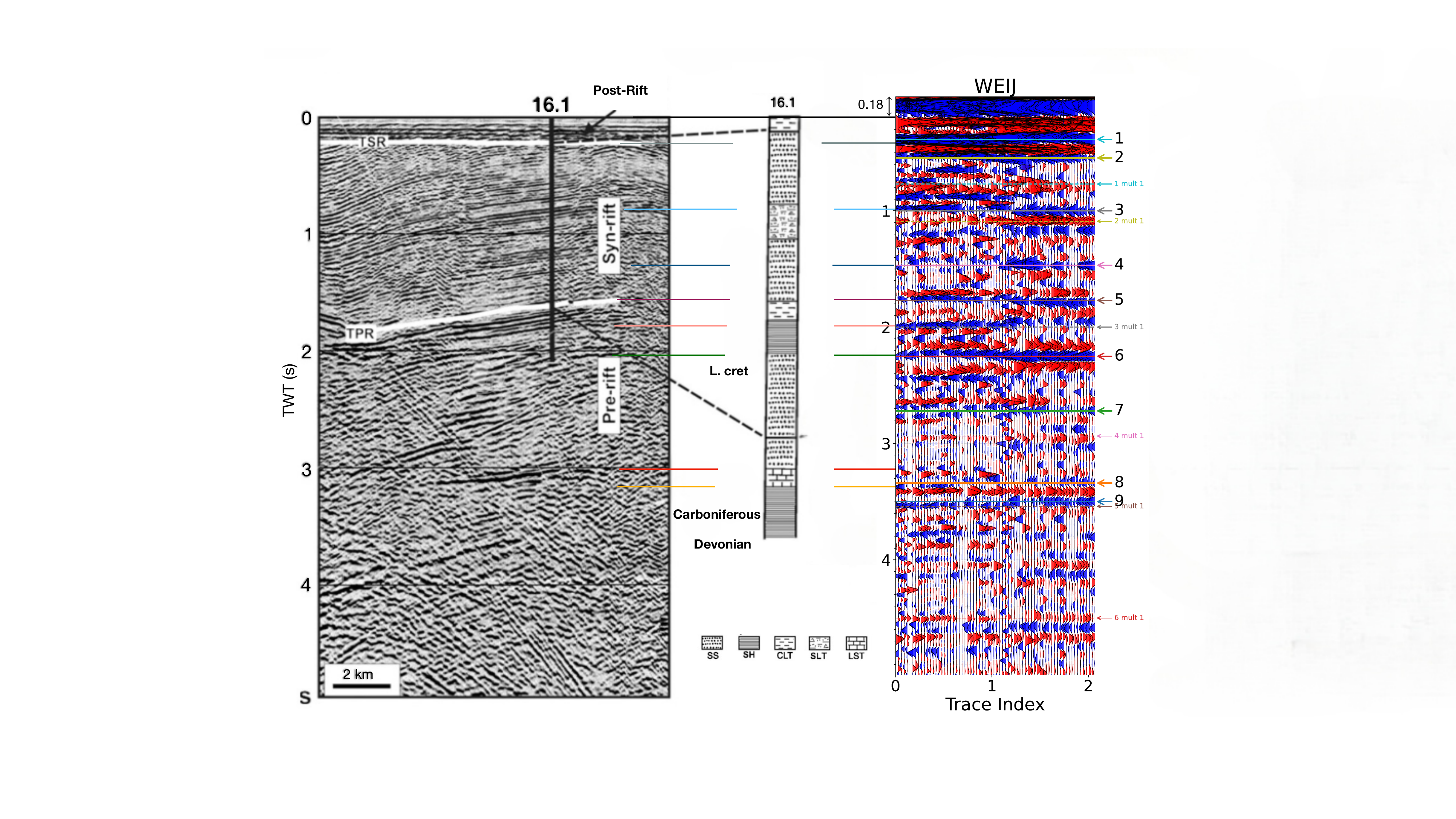}
	\end{centering}
	 \vspace{-2.2cm}
	\caption{Validation of ambient noise autocorrelation sections by comparison with active-source Multichannel Seismic (MCS) data and borehole cores drilled in the vicinity of the WEIJ station (see relative locations in Fig.~\ref{GeologyF}). The left and right sections show the first 5~s TWT of the 2D multichannel seismic (MCS) reflection profile after \citet{Antobreh2009} and the high-frequency (3–13~Hz) ambient noise autocorrelation reflection section, respectively. The central panel presents depth-domain lithological logs from the nearby exploration borehole (Well 16.1), which penetrates to ~4~km depth. The estimated depth for 5~s TWT is around 14~km. The noise section is shifted upward by 0.18~s to account for the 210~m station elevation relative to the sea-level datum of the MCS survey. The MCS section reveals three major tectono-stratigraphic regimes, including pre-rift, syn-rift, and post-rift sequences. Seismic horizons at the top of the syn-rift and within the syn-rift interval show strong correspondence between the ambient noise and MCS sections, as does the sub-syn-rift reflector. In addition, the pronounced lithological boundary between Carboniferous sediments and the underlying Devonian basement (indicated by the green line) is more clearly imaged in the ambient noise autocorrelation section than in the active-source MCS profile, highlighting the complementary resolution of passive seismic methods.}
	\label{validate}
\end{figure}

To validate the reliability of the ambient noise autocorrelation sections and the accuracy of the derived reflection response series, we conducted a comprehensive review of the available geophysical literature to identify the most consistent sets of exploration wells and reflection seismic data acquired in the study area. Previous studies have reported several 2D seismic reflection profiles offshore southern Ghana, along with a number of exploration wells drilled along the southern margin, primarily targeting offshore hydrocarbon reservoirs and investigating the regional tectonic framework \citep{delteil1974continental, Antobreh2009}. Among the available datasets, we identified a 2D \ac{MCS} reflection profile that closely overlaps the WEIJ station, together with a nearby exploration borehole (Well 16.1) located in close proximity to this station \citep{Antobreh2009} (see Fig.~\eqref{GeologyF}). The borehole reaches a depth of approximately 4~km and thus provides an independent constraint for validating the reflection interfaces inferred from the ambient noise autocorrelation analysis.

Figure~\ref{validate} compares the MCS reflection section, the borehole log, and the high-frequency (3–13 Hz) ambient noise autocorrelation–derived reflection response. Both the ambient noise reflection section and the MCS profile are displayed in the \ac{TWT} domain, whereas the borehole information is presented in the depth domain.  To ensure a consistent reference datum, the ambient noise reflection section was shifted upward by 0.18~s in the TWT domain to account for the 210~m elevation of the WEIJ station relative to the sea-level datum used in the MCS survey.
 Accordingly, the seismic reflection profile reveals three major tectono-stratigraphic regimes. The deeper part of the section corresponds to the pre-rift stage, characterized by active tectonic deformation prior to the onset of rifting in the region. The intermediate depth range represents the syn-rift phase, marked by extensional faulting and syn-tectonic sedimentation. The shallow portion of the section corresponds to the post-rift stage, characterized by relatively stable tectonic conditions and laterally continuous, sub-horizontal sedimentary sequences \citep{Antobreh2009}.

As shown in Figure~\ref{validate}, the seismic horizons between the top of the syn-rift sequence and the overlying post-rift (recent) sediments (around 0.4~s in the noise section) show an excellent match between the two time-domain sections. Similarly, several mid-range reflectors within the syn-rift sequence (0.98~s, 1.75~s, 1.95~s) are consistently identified in both datasets. The reflector associated with the sub-syn-rift unit also exhibits very good agreement between the two sections (3.2–3.5~s). Notably, the strong lithological contrast at the boundary between the Carboniferous formations and the underlying Devonian basement (2.25~s) is clearly imaged in the ambient noise autocorrelation-derived reflection section, whereas this interface is less distinctly resolved in the active-source MCS profile.

\subsection{Frequency Stability of Reflection Signals}

To assess the reliability of our reflections, we conducted a frequency stability test for the data by tracking consistent signals across different frequency ranges. We applied the workflow on 30 days of consecutive days of data for MRON station.  We mark those reflection signals that arrive at a consistent depth (or lag time) over a broad range of frequencies and critically, those that also maintain their presence at lower frequencies in the range 0.79-4.02Hz. Signals that are only stable at higher frequency bands are likely artifacts or high-frequency noise that are not related to the true reflectivity of the gross subsurface structure.

\begin{figure}
	\begin{centering}
		\hspace{-0.2cm}
		\includegraphics[width=1.\textwidth]{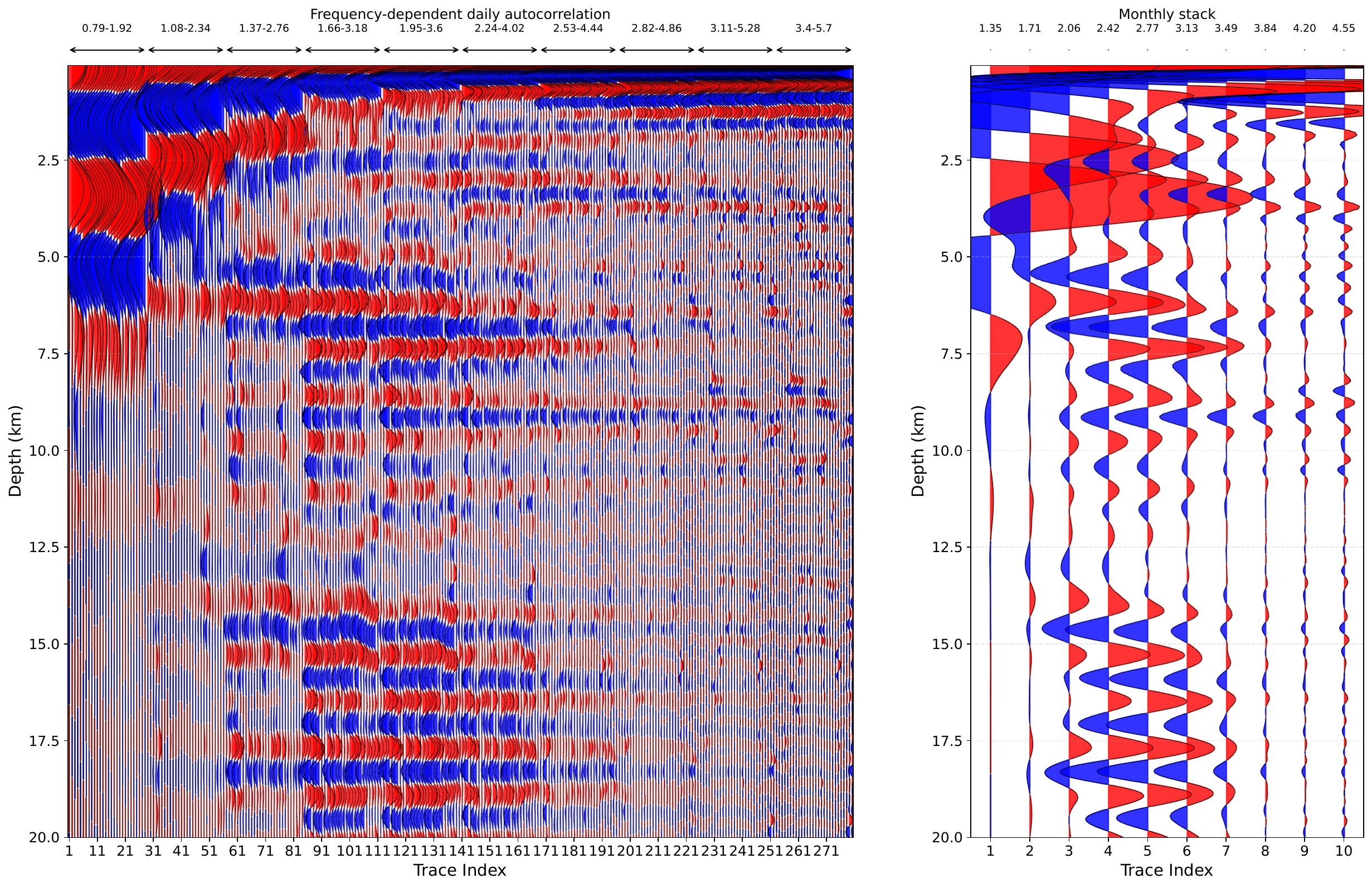}
	\end{centering}
	\vspace{-0.2cm}
	\caption{Left Panel: Frequency stability test of the time-domain reflection response for the ambient noise autocorrelation at station MRON. The section displays amplitude-recovered traces stacked across ten overlapping, progressively higher-frequency bands (0.79–1.92 Hz, 1.08–2.34 Hz, 1.37–2.76 Hz, 1.66–3.18 Hz, 1.95–3.60 Hz, 2.24–4.02 Hz, 2.53–4.44 Hz, 2.82–4.86 Hz, 3.11–5.28 Hz, and 3.40–5.70 Hz). Right Panel: Stack of daily autocorrelations for each frequency band to obtain the monthly stack. Reflection signals arriving at identical lag times across a broad range of these frequency bands are considered robust structural reflections from the upper ~14 km of the crust, whereas events appearing isolated only within the higher-frequency bands are disregarded as artifacts or unassociated high-frequency noise. The application of inverse exponential gain highlights continuity and relative reflection strength within the depth domain.}
	\label{validatef}
\end{figure}

To systematically inspect signal stability, autocorrelation sections have been calculated and stacked, and then plotted across six overlapping frequency bands (0.79–1.92 Hz, 1.08–2.34 Hz, 1.37–2.76 Hz, 1.66–3.18 Hz, 1.95–3.60 Hz, and 2.24–4.02 Hz). The result has been shown in Fig. \eqref{validatef}. Once the robust arrivals were established across the varying frequency ranges, amplitude-recovered stacks allowed us to properly identify structural boundaries within the upper 14 km of the section.

\subsection{Interpretation of the Depth domain noise sections}

The time- and depth-domain autocorrelation sections derived from the five seismic stations provide new constraints on the crustal architecture beneath southern Ghana and its eastern margin. When interpreted within the regional geological framework and integrated with available active-source seismic and well data, the results reveal a set of laterally coherent crustal interface zones that persist across these tectonic domains. These interfaces represent the inherited Paleoproterozoic structure of the Birimian crust and its systematic modification toward the Pan-African Dahomeyan belt and the Mesozoic rifted continental margin. Interpretations presented here are strictly guided by observed autocorrelation coherence and continuity, rather than assumptions of discrete lithological boundaries.

\begin{sidewaysfigure}
\centering
\includegraphics[width=\textheight]{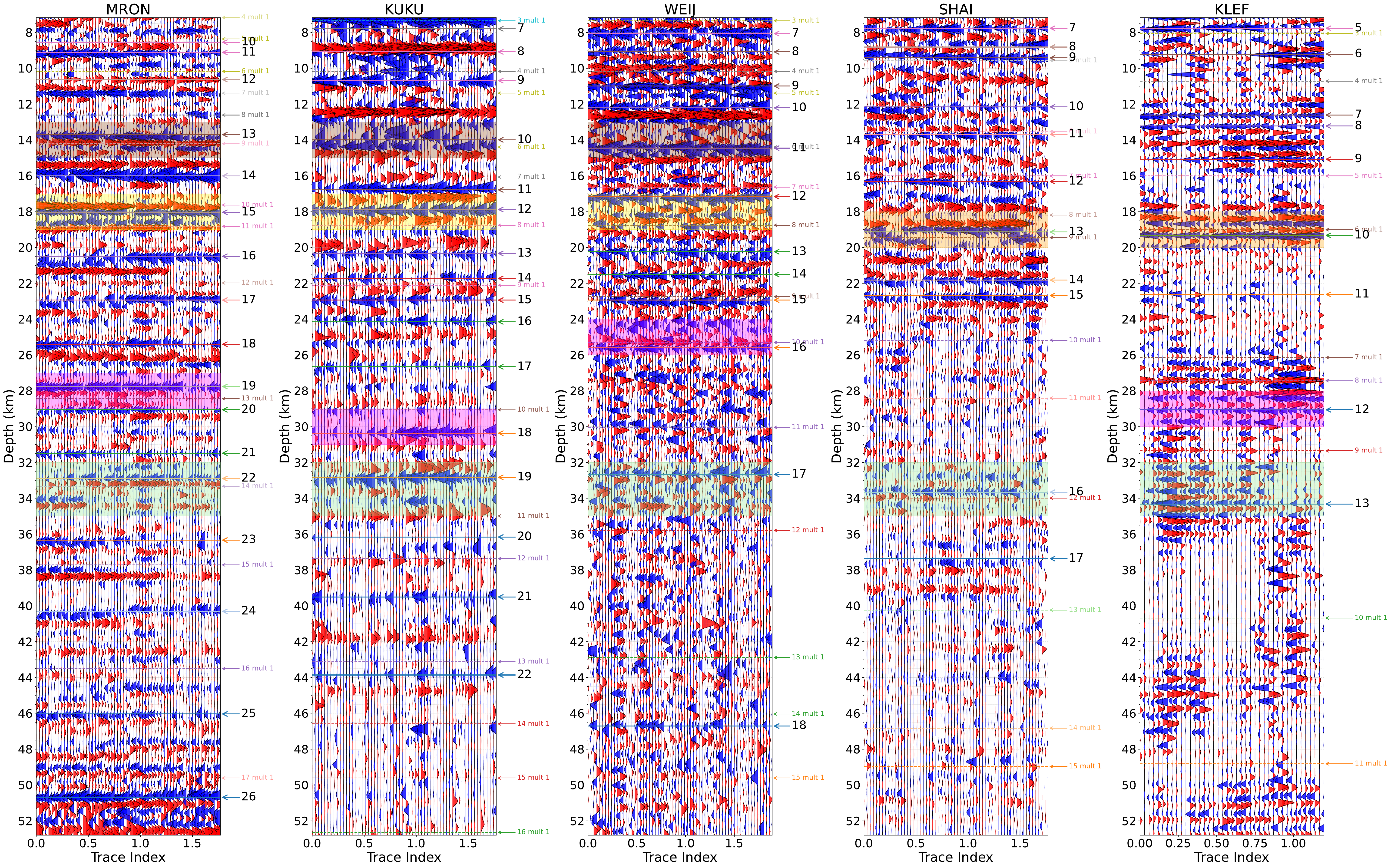}
\caption{Low-frequency reflection responses (1–6 Hz) obtained using SAANA, displayed in depth domain, for the stations KUKU, MRON, WEIJ, KLEF, and SHIAI, optimized for imaging deeper crustal structures. Coherent reflection phases appear as laterally continuous horizontal bands, indicating persistent subsurface impedance contrasts retrieved from ambient seismic noise. The enhanced reflectivity confirms the presence of key crustal interfaces beneath the station.} \label{deep_horizon_Area}
\end{sidewaysfigure}

We present geological interpretations based on the low-frequency sections (1–6~Hz) converted to the depth domain (Fig.~\ref{deep_horizon_Area}) using the velocity model calculated in Sec.~\eqref{vell}. MRON and KUKU are located within the interior of the Birimian Supergroup. Although they belong to the same geological unit, their separation distance ($\sim$167~km) is expected to result in different sequences of horizons from shallow to deep formations (Fig.~\ref{deep_horizon_Area}). In this context, WEIJ is situated at the eastern margin of the Birimian domain, adjacent to the Togo Structural Unit. Despite belonging to a different geological unit at the surface, its close proximity to the Birimian domain, together with the relatively short distance between the WEIJ and KUKU stations ($\sim$68~km), suggests that similar geological interfaces are present, as evidenced in Fig.~\ref{deep_horizon_Area}, at least below the upper crust within the depth interval of 8–24~km. Moving eastward, SHAI and KLEF (with 85km distance) are located within the Dahomeyan Complex, a Precambrian basement unit. Interestingly, the noise correlation sections of these two stations show a very good match of layers in the upper, middle, and lower crust at depths of approximately 9, 12, 14, 19, and 32~km.

A prominent and laterally coherent zone of elevated autocorrelation energy is observed between approximately 13 and 15~km depth beneath MRON, KUKU, and WEIJ (Fig.~\ref{deep_horizon_Area}), highlighted by light brown area. Rather than a single sharp reflector, this depth interval is characterized by clusters of strong and continuous autocorrelation peaks, indicating a distributed impedance contrast within the upper-crust. The consistency of this interface across stations located within the Birimian interior and at its eastern margin suggests a regional crustal unconformity inherited from Paleoproterozoic tectonic processes. The presence of an interface beneath WEIJ in the same depth indicates that Neoproterozoic deformation related to the Togo Structural Unit did not significantly disrupt the mid-crustal Birimian fabric at this depth. This observation is consistent with geological models that describe the Togo Unit as a thin-skinned thrust system, with deformation largely confined to the upper crust while deeper Paleoproterozoic structures remain preserved.


A second laterally continuous reflector is identified at depths of approximately 17–19~km beneath stations MRON, KUKU, and WEIJ, highlighted by the yellow transparent area (Fig.~\ref{deep_horizon_Area}). This horizon likely corresponds to an upper–middle crustal interface associated with a pronounced lithological contrast. A comparable reflector is also observed beneath stations SHAI and KLEF; however, it occurs at slightly greater depths, around 18–20~km. This systematic deepening beneath SHAI and KLEF may reflect the presence of thicker low-velocity sequences at these sites. When converting the time-domain sections to depth using a common near-surface velocity model for all stations, such lateral variations in shallow velocity structure can introduce apparent depth shifts in the imaged interfaces. Comparison with the depth distribution of local earthquakes recorded by the GHDSN network between 2012 and 2014 (Fig.~\ref{deep_horizon_Area}) shows that most seismicity is concentrated between 10 and 18~km depth. This correspondence suggests that the imaged reflector may approximate the brittle–ductile transition and represent the lower boundary of the active seismogenic zone in southern Ghana.

In the deeper parts of the section, in accordance with the expected stable regional geology and horizontally layered stratigraphy of the West African Craton (WAC), a consistent interface is observed in the deep part of all sections at a depth range of 32--34~km, highlighted by a transparent green area at all stations. This horizon is clearly visible at MRON, KUKU, WEIJ, and SHAI stations, and can also be identified at KLEF, despite distorted waveforms with highly illuminated side lobes of the Ricker wavelet, possibly caused by a strong reflector in the very shallow subsurface structures beneath the KLEF station. The depth range of this horizon and its lateral consistency confirms that it represents an interface between the middle and lower crust within the WAC in southern Ghana.


A deeper interface zone is observed between approximately 32 and 35~km depth beneath all stations. Although broader and less sharply defined than shallower horizons, its persistence across the study area indicates a regionally significant feature.
Regional estimates of Moho depth beneath southern Ghana place the crust--mantle boundary at approximately 41--45~km. The reflector identified at 32--35~km is therefore interpreted as an intra--lower-crustal or Moho-proximal interface rather than the Moho itself. This horizon may represent the upper portion of a thick mafic lower crust or a transitional zone approaching the crust--mantle boundary.

 We also analyzed the depth sections to extract and correlate the published crustal composition beneath the GHDSN stations in southern Ghana. In this context, \cite{akpan2016crustal} estimated the depth of the felsic–mafic transition by conducting a joint inversion of Rayleigh wave ellipticity and teleseismic receiver functions [see Table. \eqref{akpan}]. According to their study, this boundary is defined as a zone in which the S-wave velocity increases abruptly from values $<4$~km/s to $>4$~km/s.  \cite{akpan2016crustal} reported that the felsic–mafic transition is sharp within the cratonic zones, occurring at depths of 28~km beneath MRON, 30~km beneath KUKU, AKOS, and KLEF, and 25~km beneath WEIJ. Although the felsic–mafic transition is primarily characterized by a significant increase in S-wave velocity, a corresponding change in P-wave velocity is also expected due to the associated lithological contrast.  
 This behavior is evident in our noise-derived sections, where strong and distinct reflections are observed at depths of 28~km beneath MRON, 30~km beneath KUKU, 26~km beneath WEIJ, and 29~km beneath KLEF. In contrast, no sharp reflector is observed at SHAI within this depth range, consistent with the absence of a well-defined felsic–mafic transition at this site as reported by \cite{akpan2016crustal}.

\begin{table}[ht]
	\centering
	\caption{Summary of crustal composition and Felsic-Mafic transition depth beneath GHDSN stations in southern Ghana. Transition depths after Akpan et al.\ (2016).}
	\renewcommand{\arraystretch}{1.1}
	
	\resizebox{\textwidth}{!}{
		\begin{tabular}{l l p{3.2cm} p{3.6cm} p{2.8cm} p{2.8cm}}
			\hline
			\textbf{Station} & \textbf{Tectonic Domain} & \textbf{Upper Crust Composition} & \textbf{Mid-Crust Composition} & \textbf{Lower Crust Composition} & \textbf{Felsic--Mafic Transition Depth (km)} \\
			\hline
			WEIJ & Dahomeyide Belt & Felsic--intermediate gneiss/migmatite & Mixed felsic + mafic (amphibolite interlayers) & Mafic-dominant & 25 \\
			SHAI & Dahomeyide Belt & Felsic--intermediate cover/gneiss & Mafic-rich (HP mafic granulites present) & Mafic-dominant & No sharp transition \\
			AKOS & Dahomeyide Belt & Felsic--intermediate (migmatite/granitoid) & Strongly mixed felsic + mafic & Mafic & 30 \\
			KLEF & Dahomeyide Belt & Felsic (granitoid gneiss dominant) & Felsic--intermediate with mafic lenses & Mafic & 30 \\
			KUKU & Birimian Terrane & Mixed; felsic granitoids + mafic volcanics & Intermediate, layered & Mafic & 30 \\
			MRON & Interior Birimian & Intermediate--felsic (granitoid-dominant regionally) & Intermediate with mafic layering & Mafic & 28 \\
			\hline
		\end{tabular}
	} \label{akpan}
\end{table}

\subsection{Seismicity and Links to Active Tectonic Regions}

We present a revised and rigorously reprocessed seismic catalog and evaluate its implications for understanding seismicity patterns and their relationship to active tectonic regions. This dataset was previously processed by \cite{GhanaSeismicity2023,ipiml}; however, those studies employed different association (PyOcto versus REAL) and location (Hypoinverse versus Nonlinloc) methods and relied on a relatively inaccurate velocity model for both event association and hypocenter estimation. The key difference in the present study is the inclusion of the Ivory Coast station, which (1) provides western coverage for the majority of detected events and substantially reduces their azimuthal gaps, and (2) contributes a number of high-quality P and S picks to the association and location process.  This improvement leads to a significant increase in the number of associated events (740 in this study compared to 461 in \cite{ipiml} (1.61 times larger)) and allows a more robust velocity analysis by better constraining the deeper layers of the velocity model, comparing to the previously reported velocity model obtained by incorporating only 73 events \citep{GhanaSeismicity2023}. The new velocity model was obtained by inverting 3839 P and S phases associated to  552 events, each recorded with at least six phases and at a minimum of two stations with both P- and S-phase arrivals.  Moreover, the integration of the Côte d’Ivoire station significantly enhanced the azimuthal coverage of events from the western direction, providing a strong constraint on hypocentral locations.

A significant concentration of seismicity is observed in Cluster~A, located in the historically active Axim region of southern Ghana, an area known for recurrent earthquakes. The Axim zone experienced a notable earthquake in December 1636, with an estimated magnitude of approximately 5.7 \citep{amponsah2002seismic}. This cluster lies within the \ac{AsFZ}, which includes subsidiary fault systems such as the Tarkwa Fault. The spatial distribution of events follows a clear NE–SW trend, consistent with the mapped structural grain of the \ac{AsFZ}, and appears to extend offshore toward the \ac{RFZ}. This pattern indicates ongoing activity along this segment of the \ac{AsFZ}, consistent with previous studies documenting scattered but persistent seismicity in southern Ghana \citep{amponsah2002seismic}. A subcluster of events is characterized by very shallow focal depths. These events may be associated with local mining-related explosions, which are common in the Axim area due to the presence of several active gold mines \citep{hilson2006gold}. Two additional groups are observed: one within the 10–20~km depth range, confined to the upper crust, and a deeper subset concentrated near $\sim$26~km, corresponding to the lower crust.

\begin{sidewaysfigure*}
	\centering
	\includegraphics[width=\textheight]{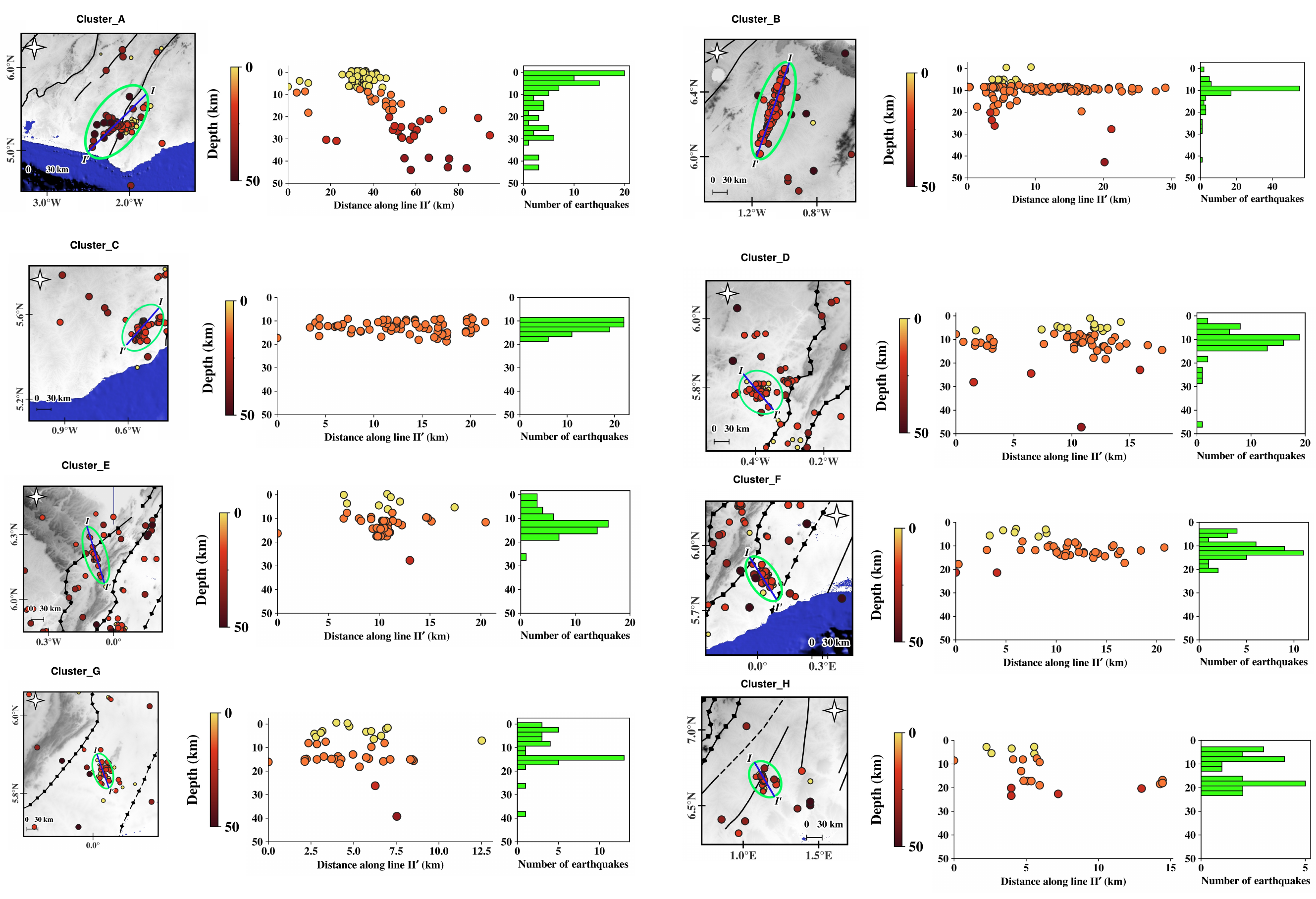}
	\vspace{-0.2cm}
	\caption{\textcolor{black}{(a) Regional seismicity map of southern Ghana and adjacent offshore areas, showing the spatial distribution of hypocenters and the locations of clusters~A–H. (b) Depth analysis of seismic clusters~A–H, with each panel presenting a localized map view of the cluster, a cross-sectional scatter plot of hypocentral depth versus distance along the designated profile line (e.g., lines~I, I$^{\prime}$), and an accompanying depth histogram (5~km bins). These panels summarize the depth structure within each cluster and highlight variations in upper-crustal seismicity across the region.}}
	\label{Final_catalog}
\end{sidewaysfigure*}


Cluster~B emerges as a newly identified zone of seismicity revealed by the improved station coverage in this study. Its alignment follows the same general structural trend observed across southern Ghana, parallel to the \ac{AkFZ} and \ac{AsFZ}. As this cluster was not detected in previous catalogs, it may represent a crustal segment that has become resolvable due to the expanded seismic network and the improved phase association approach. The depth distribution in Cluster~B exhibits a predominantly shallow character within the 5–18~km interval, peaking at 8–12~km where most events are concentrated. The location of this cluster between KUKU and MRON, together with the sharp reflectors identified in the depth-converted noise sections (Fig.~\ref{deep_horizon_Area}), suggests that the lower limit of the seismogenic zone in this part of southern Ghana is approximately 18~km. However, this interpretation should be treated with caution given the relatively limited temporal coverage of the analyzed dataset, which spans only approximately 18 months between 2012 and 2014. The location and alignment of this cluster, parallel to the main seismogenic zones and trending toward the \ac{RFZ}, may support the hypothesis of stress transfer from the \ac{MAR} through the \ac{RFZ} to inland southern Ghana, while maintaining the overall upper-crustal seismogenic regime.

Clusters~C-G as shown in Fig. \ref{Final_catalog}, together with a sparse set of earthquakes recorded along the \ac{AkFZ} toward the inland mainland, delineate a dense network of active faults within this historically and recently active seismogenic zone \citep{amponsah2002seismic}. Although some of these clusters show local deviations from the dominant NE–SW structural trend of the \ac{AkFZ}, their overall spatial distribution is consistent with this orientation, and they exhibit a highly correlated seismogenic depth range. More specifically, the depth intervals for Clusters~C (8–18~km), D (8–16~km), E (8–18~km), F (8–16~km), and G (0–18~km), together with their proximity to stations KUKU, AKOS, SHAI, and WEIJ, and considering the sharp transition zones observed in the depth sections beneath these stations, can further support the hypothesis of a brittle–ductile transition at approximately 18~km depth in this region. This major structural feature, confirming that seismicity in this region follows the known tectonic framework. The extension of activity to the northeast of the network shows that seismicity continues inland beyond the areas covered by previous catalogs. Clusters~C and D display a mid- to upper-crustal mode between 5--16~km, with tightly grouped hypocenters (rewrite c and D based on the new figure). This clustering is consistent with activity on moderately dipping upper-crustal faults within the brittle lithosphere. 

Cluster~E exhibits a broadened depth distribution, with few events at shallow depths (0--5~km) and a main concentration between 6--19~km. It also includes deeper outliers at 19--30~km, with scattered hypocenters extending to approximately 47~km, possibly reflecting multi-segment fault geometry or variable lithospheric rheology along the \ac{AkFZ}. Cluster~F mirrors Cluster~C, with most hypocenters between 5--17~km and scattered events extending to 30~km, suggesting consistent upper-crustal strain accumulation. Cluster~G shows a main concentration of events between 5--19~km, with subtle deeper tails reaching 25--48~km. This pattern may indicate downdip extension associated with stress transferred from the \ac{RFZ}. Cluster H lies outside the Ghana network but remains well constrained due to improved station geometry. It aligns with a mapped structural trend in southwestern Togo and follows the NE–SW orientation common across the region. The clear clustering suggests localized deformation in this area, consistent with regional tectonic patterns described in earlier studies of West Africa. Seismicity in Cluster H is primarily shallow (0–10 km), with scattered events extending to approximately 24 km, suggesting diffusive seismicity near the strain front.

A key outcome of this study is the absence of significant seismicity along the purported \ac{CBF}, despite improved detection capabilities and the contribution of the Ivorian station. The lack of activity during the 2012–2014 period, consistent with earlier findings \citep{amponsah2002seismic,kutu2013seismic}, casts doubt on the activity or even the existence of this proposed fault. These results provide new evidence that the \ac{CBF} is likely inactive or does not represent a major tectonic structure. Our catalog, recording 741 earthquakes over a period of just 18 months, reveals a high concentration of clustering along localized active seismogenic zones. This dataset provides specific insights into Ghana’s seismicity which are crucial for understanding regional geotectonic processes and fault activity, and for guiding local seismic hazard assessments in Ghana and surrounding regions \citep{tavakolizadeh2022extent, tavakolizadeh2024faultquake, tavakolizadeh2025fqsha}.

\section{Conclusion and Future Works} \label{conclusionn}
This study presents a comprehensive analysis of the crustal architecture and seismotectonics of southern Ghana using single-station ambient noise autocorrelation (SSANA) applied to the Ghana Digital Seismic Network (GHDSN) dataset recorded during 2012--2014. Noise sections were compiled from multiple stations across two frequency ranges: 1--6~Hz for imaging deep crustal structures and 3--13~Hz for characterizing shallow structures.

To overcome the intrinsic limitations of one-dimensional zero-offset measurements, an optimal regional velocity model was derived through joint grid-search inversion of local earthquake arrival times. To this end, continuous waveform data from the GHDSN and an additional station in Côte d'Ivoire were reprocessed for the detection and association of P- and S-wave phases of local earthquakes. The detected phases were integrated into a joint inversion scheme to simultaneously estimate an accurate 1-D velocity model for the region. This velocity model served a dual purpose: converting two-way travel-time (TWT) noise sections to the depth domain, and linking accurately located earthquakes to the active tectonics of the region and to reflective horizons identified in the noise sections.

The application of SSANA, employing Phase Cross-Correlation (PCC) and Phase-Weighted Stacking (PWS), enabled the robust retrieval of zero-offset P-wave reflection responses beneath each station. By incorporating an additional station in Côte d'Ivoire, the high azimuthal gap for major events in southern Ghana were substantially mitigated, improving both azimuthal coverage and the ray coverage to deeper crustal layer.

The validity of the velocity model and autocorrelation results was confirmed through comparisons with independent multichannel seismic (MCS) profiles and borehole data. At station WEIJ, in particular, lithological boundaries such as the Carboniferous--Devonian contact were clearly resolved.

The reflectivity sections, converted to depth using the optimized 1-D velocity model, reveal laterally coherent impedance contrasts that correlate well with known geological features. Key crustal interfaces were identified, including sediment-basement boundaries within the Voltaian Basin and a prominent mid-crustal discontinuity at depths of 13--15~km, interpreted as a Paleoproterozoic unconformity preserved across the Birimian and marginal terranes. The depth sections also clearly image the felsic--mafic boundary at several stations, with an identified transition depth of $\sim$26--30~km that aligns strongly with independent shear-wave velocity models derived from joint inversions of Rayleigh wave ellipticity and teleseismic receiver functions \citep{akpan2016crustal}, thereby validating the compositional stratification of the crust.
Our improved 1-D P-wave velocity model differs significantly from previous regional estimates \citep{custodio2022seismicity}, being characterized by a very low-velocity surface layer in the uppermost 1~km, relatively lower velocities extending to mid-crustal depths (4.90--6.20~km/s), and a distinct velocity jump in the lower crust near 18~km depth.

The integration of the new velocity model facilitated the compilation of a high-resolution earthquake catalog for the period 2012--2014. The re-analysis yielded 741 accurately located events---a 1.6-fold increase over previous studies \citep{ipiml}---benefiting substantially from the improved azimuthal coverage afforded by the additional station in Côte d'Ivoire. Spatial analysis delineates eight distinct clusters (A--H) aligned with the Akwapim and Ashanti Fault Zones, while the Coastal Boundary Fault (CBF) remained notably quiescent throughout the observation period.

A critical finding of this work is the constraint on the seismogenic depth, estimated at approximately 18~km based on depth histograms. This depth correlates strongly with sharp, high-amplitude reflections observed in the depth-domain ambient noise sections, suggesting that these reflectivity contrasts delineate the regional brittle--ductile transition. Furthermore, a meaningful tectonic link between the Akwapim Fault Zone (AkFZ) and the Romanche Fracture Zone (RFZ) is identified, confirming stress transfer from the RFZ to the active seismic zones of southern Ghana. The seismic quiescence of the CBF during the observation period suggests it may currently be inactive or locked.

Collectively, these results provide a refined crustal framework for the southern margin of the West African Craton, demonstrating the efficacy of passive seismic interferometry for characterizing detailed crustal structure in intraplate settings. The derived velocity and structural models offer critical constraints for future seismic hazard assessments and resource exploration in the region.

\section*{Conflict of Interest declaration}
The authors declare there are no conflicts of interest for this manuscript.

\acknowledgments

\section*{Data Availability Statement}

he data files used in this paper are available at \citep{gh} and  \citep{https://doi.org/10.7914/sn/gt}. All waveform data were obtained from this source.

\begingroup
\hyphenpenalty=10000
\exhyphenpenalty=10000
\bibliography{references}
\endgroup
\begin{acronym} 

\acro{WWSSN}{World Wide Standard Seismograph Network}
\acro{GSD}{Geological Survey Department}
\acro{AR}{accuracy rate}
\acro{AFZ}{Akwapim Fault Zone}
\acro{GDSN}{Ghana Digital Seismic Network}
\acro{CCGN}{cross-correlation geometrical normalized}

\acro{PCC}{phase cross-correlation geometrical}

\acro{tfPWS}{time-frequency Phase Weighted Stacking}

\acro{PWS}{Phase Weighted Stacking}
\acro{MCS}{Multi Channel Seismic}

\acro{WAC}{West African Craton}

\acro{PGA}{peak ground acceleration}
\acro{DSHA}{deterministic seismic hazard assessments}

\acro{ACFs}{autocorrelation functions}
\acro{EGF}{empirical Green’s function}

\acro{GTSN}{Global Telemetered Seismograph Network}

\acro{TWT}{two-way travel times}
\acro{GTSN}{Global Telemetered Seismograph Network (USAF/USGS)}

\acro{GHDSN}{Ghana Digital Seismic Network}
\acro{HRSCCW}{High Resolution Seismic Catalog Compilation Workflow}

\acro{PCC}{Phase Cross-Correlation}

\acro{SSANA}{single station ambient noise autocorrelation}
\acro{EQT}{EQTransformer}
\acro{S-EQT}{S-EQTransformer}

\acro{WAC}{West African Craton}
\acro{MAR}{Mid-Atlantic Ridge}
\acro{GOG}{Gulf of Guinea}
\acro{TPR}{True Positive Rate}
\acro{FNR}{False Negative Rate}
\acro{FPR}{False Positive Rate}
\acro{TNR}{True Negative Rate}
\acro{CFZ}{True Negative Rate}
\acro{CFZ}{Chain Fractured Zone}
\acro{RMS}{Root Mean Square error}
\acro{Mc}{Magnitude of Completeness}

\acro{SPFZ}{St. Paul’s Fracture Zone}

\acro{AkFZ}{Akwapim Fault zone}

\acro{CS}{Conservative Strategy}

\acro{SNR}{signal-to-noise ratio}

\acro{WAP}{West African Plate}
\acro{STEAD}{STanford EArthquake Dataset}
\acro{DL}{Deep Learning}
\acro{AI}{Artificial Intelligence}
\acro{HAM}{Hierarchical Attention Mechanism}
\acro{ISC}{International Seismological Center}
\acro{USGS}{United States Geological Survey}
\acro{GHGS}{Ghana Geological Survey}
\acro{GHDSN}{Ghana Digital Seismic Network}

\acro{AsFZ}{Ashanti Fault Zone}
\acro{DL}{Deep Learning}

\acro{IRIS}{Incorporated Research Institutions for Seismology}
\acro{CBF}{Coastal Boundary Fault}
\acro{RFZ}{Romanche Fracture Zone}
\end{acronym} 

\end{document}